\documentclass[aps,pre, amsmath,amssymb, reprint,]{revtex4-2}

\usepackage{graphicx}
\usepackage{dcolumn}
\usepackage{bm}
\usepackage{amsfonts,amsthm,mathbbol}
\usepackage[utf8]{inputenc}
\usepackage[T1]{fontenc}
\usepackage{mathptmx}
\usepackage{soul}
\usepackage[normalem]{ulem}
\DeclareSymbolFont{greekletters}{OML}{cmr}{m}{it}
\DeclareMathSymbol{\varrho}{\mathord}{greekletters}{"25}
\usepackage{color}
\usepackage{tabularx}
\usepackage{multirow}
\usepackage{tabularray}
\usepackage[colorlinks=true, linkcolor=blue, citecolor=blue, urlcolor=blue]{hyperref}
\usepackage[normalem]{ulem}
\usepackage{cancel}
\usepackage{twemojis}
\usepackage{tikzsymbols}
\usepackage{mathtools}
\DeclarePairedDelimiter\bra{\langle}{\rvert}
\DeclarePairedDelimiter\ket{\lvert}{\rangle}
\DeclarePairedDelimiterX\braket[2]{\langle}{\rangle}{#1\,\delimsize\vert\,\mathopen{}#2}
\DeclarePairedDelimiterX\ketbra[2]{\delimsize\vert}{\delimsize\vert}{#1\,\rangle\,\langle\,#2}

\newcolumntype{Y}{>{\centering\arraybackslash}X}

\newcommand{\hp}{\hat{p}}
\newcommand{\hq}{\hat{q}}
\newcommand{\hr}{\hat{\rho}}
\newcommand{\cc}{\hat{\textbf{a}}}
\newcommand{\cd}{\hat{\textbf{a}}^{\dag}}

\definecolor{red}{rgb}{1,0,0}

\begin{document}

\title{Quantum Thermodynamics of Electron Transport along Chains of Redox Centers}

\author{Fabrizio Cleri}
\email[Corresponding author: ]{fabrizio.cleri@univ-lille.fr}  
\affiliation{University of Lille, Institut d'{\'E}lectronique, de Micro{\'e}lectronique et de Nanotechnologie (IEMN CNRS UMR8520) and Departement de Physique, F-59652 Villeneuve d'Ascq, France }
\altaffiliation[Presently on leave at: ]{Department of Chemical Systems Engineering, University of Tokyo, Tokyo 113-8656, Japan}

\author{Ralf Blossey}
\email{ralf.blossey@univ-lille.fr}
\affiliation{University of Lille, Unit{\'e} de Glycobiologie Structurale et Fonctionnelle (UGSF CNRS UMR8576), F-59000 Lille, France}

\author{Stefano Giordano}%
\email{stefano.giordano@univ-lille.fr}  
\affiliation{University of Lille, CNRS, Centrale Lille, Univ. Polytechnique Hauts-de-France, UMR 8520 - IEMN - Institut d'{\'E}lectronique, de Micro{\'e}lectronique et de Nanotechnologie, F-59000 Lille, France}

\date{\today}

\begin{abstract}
Intramolecular electron transport in biological systems is typically described as a diffusive hopping process, according to the semi-classical rate theories of Marcus and Hopfield combined with classical Pauli-type master equations. However, the possibility that non-trivial quantum mechanical effects could play a functional role in the transport dynamics in certain biomolecular processes has attracted increasing attention. Here, we extend the quantum mechanical model of open system dynamics by the Lindblad equation to a key biological component, the long chains of redox centers based on iron-sulfur clusters or heme groups that are widespread in many biological organisms, where they realize the cellular respiration. This approach allows to explore a wide range of physical parameters, showing key features of electron transport in these multi-domain protein structures. We pay particular attention to heat and entropy transfer between the electrons and the protein bath, which constitutes a benchmark of physical realism for the models. Electron currents, average transfer times and relative efficiency of the transport process are also explicitly characterized.
\end{abstract}

\maketitle

\section{Introduction}

\begin{figure}[t]
\centering
\includegraphics[width=\columnwidth]{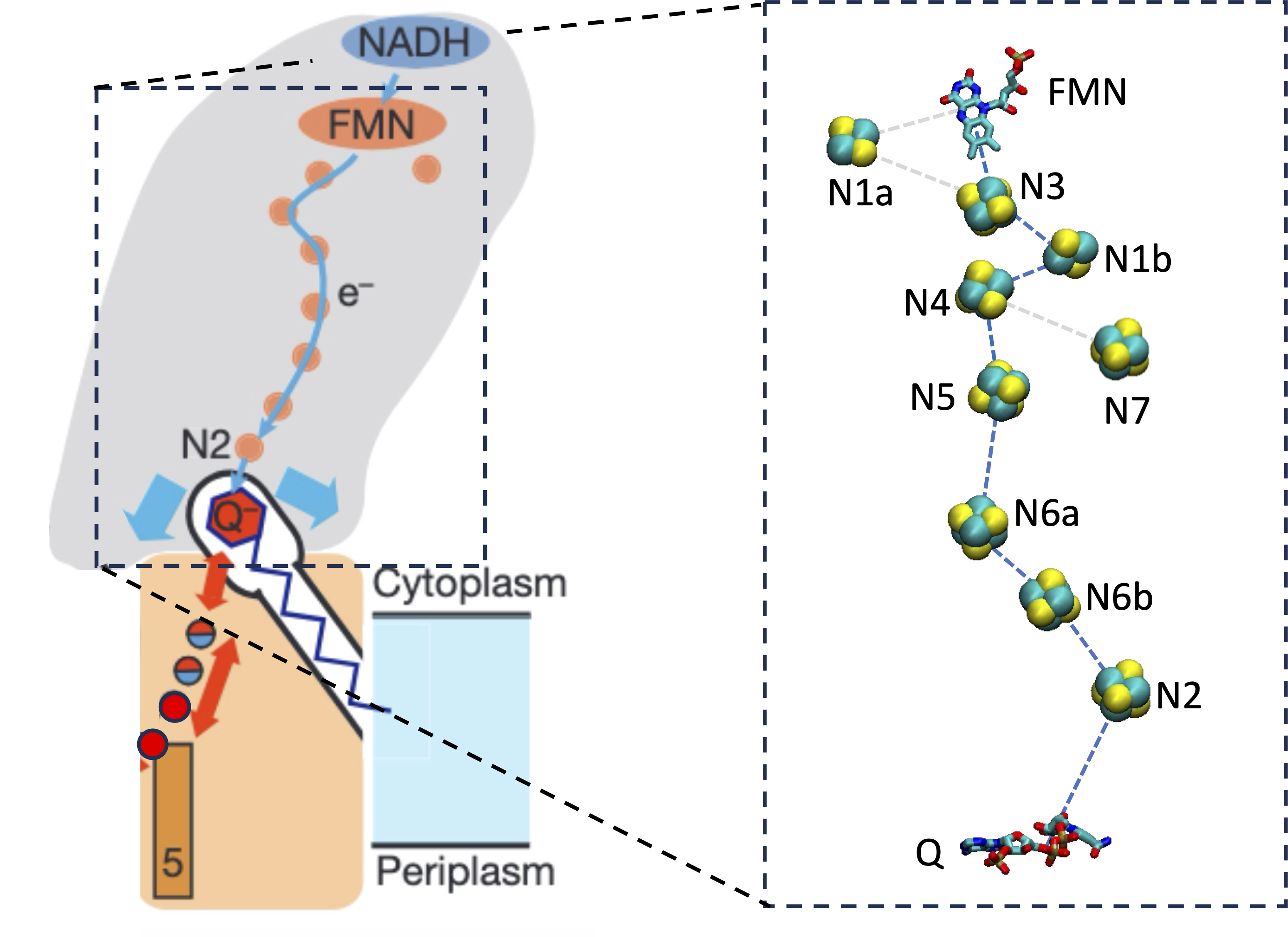}
\caption{Schematic of the mitochondrial Complex-I. Left: the peripheral arm (grey) containing the redox chain initiated by the NADH oxidation, continues to the integral arm of the complex (orange, only the first subunit of seven is shown). The final reduction of quinone (Q) induces a sequence of mechanical transformations that allow the passage of four H$^+$ in the periplasmic space. Right: the molecular arrangement (from PDB 7B93 \cite{musmus}) of the Fe-S clusters between the top flavin mononucleotide and the terminal quinone.}
\label{figc}
\end{figure}

The molecular mechanisms of respiration are remarkably similar across widely different organisms. Respiration is ultimately a process of extracting electrons from the supply (NADH) and transporting them to the final acceptor oxygen, generating proton motive force to recharge ATP reserves. As a general rule, a complex of transmembrane proteins contains a series of redox centers, that allow electrons to enter the mitochondria (in eukaryotes) or the plasmic membrane (in prokaryotes), by a successive chain of charge transfers in which each center is alternately reduced and oxidized. All eukaryotes, from fungi to animals with little exceptions, use minor variants of the Complex-I-III electron transport chain \cite{nicklane,shin26}. In mitochondrial Complex-I, electrons travel approximately 90 \AA~via quantum mechanical tunneling. Pairs of electrons are injected from NADH to a chain of closely spaced redox centers (cofactors): the Flavin Mononucleotide (FMN) and several iron-sulfur (Fe/S) clusters (7 in human, up to 8-9 in other species) 
N1a and N1b are binuclear ([2Fe-2S], and the next N3, N4, N5, N6a and N6b are tetranuclear ([4Fe-4S]), like the terminal N2. Finally, electrons are exported from cluster N2 to ubiquinone, which accepts them at roughly -70 mV to +30 mV, driving the conformational changes required for proton pumping in the membrane domain \cite{efresaz,parey}.
The molecular structure of the hydrophilic domain of respiratory Complex I (about half the molar mass of the total complex) was solved at 3.3 \AA~resolution by Sazanov and Hindhcliffe \cite{sazza}, yielding the clean architecture of a dense packing of eight subdomains containing nine Fe/S clusters (Figure \ref{figc}). 

Compared to the remarkably conserved structure of electron transport chain in eukaryotes, the world of prokaryotes, bacteria and archaea, is a vast laboratory of electrical experimentation, with coaxial cables sprouting from membranes and an impressive diversity of molecular solutions to the same problem \cite{lynch,granick,wang25}. 
Some bacteria actually breathe iron or other metals from solid rocks. \textit{Geobacter sulfurreducens} \cite{geobact} extracts electrons from carbon sources and transfers them by a dense newtork of interacting cytochromes, onto insoluble iron oxides external to the cell.
In \textit{Shewanella oneidensis} the CymA protein hands the electrons to the MtrA/B/C porins pathway, literally a molecular extension cord that drags the electrons out of the inner membrane and drops them onto the external rock surface. Some other bacteria use the same process chain but in reverse, \textit{Acidithiobacillus ferrooxidans} imports electrons from dissolved iron Fe$^{2+}$ (rust precursor, \textit{A. ferrooxidans} lives only is extremely acidic pools at pH 1-3, where full iron oxidation is blocked), which flow inward through periplasmic carriers like rusticyanin, and enter the quinone pool to drive proton movement, effectively running the mitochondrial-style engine backward \cite{fengli}.
"Cable" bacterial colonies like \textit{Cand. electrotrix} and \textit{Cand. electronema} form shared transport chains centimeters long, across thousands of cells, to reach the colony surface where oxygen captures the electrons and the proton-motive force is distributed to the whole colony.  

Such chains of redox centers, ranging from small Fe/S clusters of a few atoms, to large cytochrome heme of >600 Da mass, depending on the detailed transport kinetics, can be very long, up 10 sites and more, with spacings ranging from a few \AA~to about 14-16 \AA. Redox potentials also vary enormously both in range, from just a few up to hundreds of mV, and in shape, from plateau with large jumps, to smoothly linear, to sawtooth.
Macroscopic quantum effects have often been evoked in such systems, e.g., temperature-independence of oxidation rate in \textit{Chromatium} \cite{devault}, spin-noise cooperation and wavepacket-like electron dynamics in Complex-I \cite{hongbao,matyushov}, electron transport in \textit{S. oneidensis} appearing to involve a mix of hopping and wave-like diffusion \cite{subra18} as well as spin-tuning \cite{mishra19}.

In these common transport systems, the electron wavefunction penetrates by quantum tunneling the classically-forbidden potential barriers, allowing fast, directional transport to the terminal electron sink. The wave nature of the electrons produces both constructive and destructive interference, when multiple tunneling pathways exist simultaneously through the protein.
The Fe/S clusters are separated by distances of 8 \AA~to 14 \AA, and act as a quantum "wire". They are positioned corner-to-corner with specific Cysteine ligands oriented toward each other, with the result of maximizing electronic coupling \cite{page}.
Specific peptide residues may bridge the gap between neighboring Fe/S clusters, serving as virtual intermediate states to greatly enhance the tunneling rate.
Moreover, hydrogen-bonding networks formed by localized internal water molecules provide alternative, low-activation-energy pathways for the electron. These may create shorter through-space, or "super-exchange" jumps (about 1.7--2.2~\AA), thereby enhancing the overall tunneling efficiency \cite{stuche1,stuche2}.

Modeling  energy and charge dynamics in molecular systems can be done with Stochastic Quantum Liouville equations (SQLE), which represent dissipation indirectly via the fluctuations of a random potential over time, allowing memory effects of the environment on the transport phenomena to be studied \cite{breuer,weiss}. By contrast, the structurally simpler GSKL or Lindblad equation describes dissipation in a direct and deterministic way, assuming that the environment is \textit{memoryless}, and that the system loses energy and coherence via practically instantaneous quantum jumps (see e.g. the recent review \cite{campaioli}). Recent work by Barford \cite{barford} compared SQLE to GSKL for an idealized case of Frenkel exciton dynamics in one-dimensional molecular systems such as J-aggregates or conjugated polymers, showing that decoherence in charge transport systems, at least in the high temperature regime (of interest in biophysics) can be effectively mapped not only as the effect of a continuous "disturbing" thermal bath (SQLE), but as a sequence of local collapse events (GSKL) that periodically reset the carrier's coherent memory. This is very good news, since the implementation of GSKL is vastly simpler and more efficient than a SQLE.

A chain of redox centers may be formally modeled using a tight-binding Hamiltonian in the Anderson-Holstein multi-site style, treating the cofactors as localized electronic sites that are quantum-mechanically coupled to each other, and to the thermally fluctuating protein environment. This method has been already used in the pioneering work by Rebentrost et al. \cite{reben}
using a Lindblad equation on photosynthetic redox centers. Similar approaches but with different scope were used by Denton et al. \cite{denton}
to model birds' magnetoreception and radical pairs in cryptochrome with a Nakajima-Zwanzig equation, or by Slocombe et al. \cite{sloco} 
to study proton tunneling in DNA with a Caldeira-Leggett equation.

The tight-binding formalism with Lindblad may be especially adapted to Fe/S and heme redox chains if conditions of weak coupling exist, characterizing the fast time scales of the electron transfer as closely approximating a Markovian process. 
The Lindblad approach typically used in studies of open quantum systems such as quantum wires, quantum optics, qubits \cite{daley,campaioli,mccauley,rudner,clerifr}, has the unique ability to guarantee the preservation of physical reality (positivity, trace conservation, and hermiticity of the density matrix) while remaining computationally simple and analytical, compared to more complex, non-Markovian treatments for strong coupling regime. Biological transport chains, however, rarely happen in the weak coupling regime, so this will be one of the main points of discussion. Another drawback of such a highly flexible approach in complex biophysical systems, may be that multiple different choices of jump operators can yield similar-looking decay rates. Without a microscopically-grounded derivation, choosing operators becomes an exercise in parameter fitting rather than fundamental physics, with the accompanying risk of violating thermodynamic consistency (Second Principle, Gibbs thermal equilibrium, non-adiabatic regime).

In the following, we will first establish the theoretical basis for the application of the weak-coupling model of open quantum systems to the multicenter redox chains, starting with the ordinary "temperature-less" version of the Lindblad dissipator, and then by introducing corrected sets of operators that satisfy detailed balance at finite temperature.  
Section \ref{totham} presents the general model system in a tight-binding approximation for the redox centers, for which the basic Lindblad formalism will be developed in the next section \ref{lindtb}. Section \ref{lindterm} will then formulate the problem of thermodynamic consistency, with two approaches to the Lindblad equation: a microscopically derived formalism grounded on the quantization of the Klein-Kramers equation, and a phenomenological approach including detailed thermal balance on the basis of the polaron expansion. Finally, some numerical solution examples will be presented in Section \ref{risulta}, to compare the three approaches and identify the peculiarities of electron transport in redox chains. In conclusion, rather than a mere comparison between different numerical methods, this work will compare three ontologies of electron transport: localized particle hopping between redox centers, vs. 
electron as a dissipative quantum probability fluid interacting continuously with the bath, vs.
electron as a thermally dressed quasiparticle, whose dynamics remain site-resolved but inherit the correct thermodynamic structure.


\section{The Total System Hamiltonian}\label{totham}

Classical treatments of electron transport usually start from the Anderson-Holstein model, originally describing electrons in a periodic lattice in which thermally excited phonons represent quantized vibrational modes, interacting with electrons in a linear approximation (see e.g. \cite{leijnse,ciuchi}).
The complete Hamiltonian $\hat{H}_{tot}$ is partitioned in three main components:
\begin{equation}
\hat{H}_{tot} = \hat{H}_{el} + \hat{H}_{nucl} + \hat{H}_{coupl}
\label{hamil}
\end{equation}

This model can be extended to a topologically linear chain of centers (redox cofactors), the "phonons" being identified with vibrational modes of the protein systems. However, in most practical cases of biophysical relevance, the transport regime is strong coupling because of the substantial protein rearrangement induced by oxidation/reduction events at the passage of electrons. 
Trying to retrieve a weak coupling regime, electrons and local vibronic modes of the proteins have been associated in a compact description by the concept of the \textit{polaron} (\cite{dorner}, see below Section \ref{lindtherm}), a bosonic quasiparticle that can greatly simplify the description of the interaction between quantum phase coherence and thermal dissipation, while retaining dynamical "dressing" effects of the final states (the electron slows down, dragging the lattice distortion with it). In this respect, a Lindblad-like description of the coupling between electron and thermal bath, in which all the components of the density matrix for localized states appear explicitly, may be easier to interpret both in electronic and thermodynamic terms, provided the interactions remain in the weak-coupling domain of applicability of the approximation.
In a tight-binding approximation, the electronic part will be described explicitly, while the nuclear/coupling part is subsumed in the thermal noise, described in the Lindblad dissipator.

The electronic Hamiltonian captures the localized energy levels of the electron on each cofactor and the quantum mechanical tunneling matrix elements between them:
\begin{equation}
\hat{H}_{el} = \sum_{i}  \varepsilon_i \ket{i} \bra{i} + \sum_{i \neq j } T_{ij} \left( \ketbra{i}{j} + \ketbra{j}{i} \right)
\end{equation}
The $\varepsilon _{i}$ are the on-site energies, localized potential energy of an electron on cofactor $i$ corresponding directly to the reduction potentials of each individual atom cluster in the protein;
the $T_{ij}$ are tunneling matrix elements, that is the electronic coupling or "hopping" amplitude between cofactor $i$ and cofactor $j$. This term drops exponentially with intersite distance $R_{ij}$ and depends heavily on the specific protein (super-)exchange pathways bridging the spots; 
the $\ketbra{i}{i}$ are the projection operators representing the electron occupying specific site $i$.

In the literature, values of on-site energy (often expressed as intrinsic standard potentials) for different redox complexes vary largely, depending on whether macroscopic experimental data or computational models of electrostatics (e.g. macrostates and Coulomb couplings) are considered \cite{couch,muhlb}.
Classical values obtained by spectroscopic (electron paramagnetic resonance, EPR) and potentiometric titration  
reflect macroscopic release/injection potentials, which include shielding effects and cooperative interaction between neighboring sites.   
For quantum dynamics or continuous quantitative electrostatic calculations (such as models solving the Poisson–Boltzmann equation or Holstein/Marcus Hamiltonian models), instead, pure on-site energies are extracted after cleaning the potential from mutual Coulomb interactions \cite{kanda,dorner,dorner2}.

The tight-binding off-diagonal couplings $T_{ij}$ (or jump integrals) are much smaller than the diagonal energies (on-site energies, even compared to Marcus' reorganization energy $\lambda$).
While on-site energies vary on scales of tens or hundreds of meV and $\lambda \sim$ 200-400 meV, the electron coupling between distant Fe-S clusters collapses exponentially due to the insulating barrier of the protein matrix.
The biophysical literature and quantum mechanical calculations (QM/MM) offer rather precise references (see e.g. \cite{moser,dorner}). 
Taking as an example the 8 cofactors in the main "quantum wire" of Complex-I, the electronic part can be visualized as an 8$\times$8 Hermitian matrix:

\begin{equation}
\hat{H}_{el} = \left( \begin{matrix} 
\epsilon _{\text{FMN}} & T_{1 2} & 0 & \dots & 0 \\ 
T_{2 1} & \epsilon _{\text{N3}} & T_{2 3} & \dots & 0 \\ 
0 & T_{3 2} & \epsilon _{\text{N1b}} & \dots & 0 \\ 
\vdots & \vdots & \vdots & \ddots & \vdots \\ 
0 & 0 & 0 & \dots & \epsilon _{\text{N2}}
\end{matrix} \right)
\end{equation}

Here the diagonal elements $\varepsilon _{i}$ control the directionality (downward energy cascade) toward the terminal N2 cluster. The off-diagonal elements $T_{ij}$ dictate the relative efficiency of coherent quantum hopping. Because all the $T_{ij}$ decay rapidly over long distances, they are essentially zero for non-adjacent clusters, making the matrix highly tridiagonal ideally suited to the tight-binding representation. However, non-zero elements such as $T_{i,i\pm 2}$ can appear in the case of super-exchange processes in which an adjacent site is skipped and electrons tunnel directly between second-neighbors (see section \ref{superex} below).

\section{ The Lindblad Formulation for Tight-Binding Centers }\label{lindtb}

In the open quantum system framework, the Von Neumann equation for the total density matrix $\hr_{tot}(t)$ of the (system + bath) in the Dirac interaction representation is given as:
\begin{equation}
\frac{d\hr _{tot}(t)}{dt}=-\frac{i}{\hbar}[\hat{H}_{tot}(t),\rho _{tot}(t)]
\end{equation}
For weak coupling, it is customary to make the iterative expansion to the second perturbative order; taking the partial trace on the degrees of freedom of the phonon bath $\hr(t) = \text{Tr}_B [\hr_{tot}(t)]$, gives the Redfield equation \cite{breuer}. 
The latter is characteristically defined by an integral of the
double commutator $[\hat{H},[\hat{H},\hr_{tot}]]$ over a finite time $\tau$, describing the memory effect of the perturbed electron dynamics.

If the phonon bath is large and at equilibrium,   
the total density matrix can be factored  as $\hr_{tot}(t) \approx \hr(t) \otimes \hr_B$, isolating the electron part $\hr$ from the thermal bath $\hr_B = \exp(-\beta H_B) / Z$.  
If, moreover, the correlation functions of the phonon bath $\langle B_i(t) B_j(t') \rangle_B$ 
decay over a time scale extremely fast with respect to the time scale of evolution of the electronic system, the regime of weak coupling is obtained. One can then obtain a purely Markovian memoryless dynamics, leading to a master equation scheme \cite{breuer,ziman,jeske}. The Gorini-Kossakowski-Sudarshan-Lindblad (GKSL) equation describes the evolution of the electron density matrix $\hr(t)$ alone, in which thermalized phonons act as a thermal bath and are traced away, generating dissipation and quantum decoherence:
\begin{equation}
\frac{d \hr(t)}{dt} = - \frac{i}{\hbar} \left[ \hat{H}_{el} , \hr(t) \right] + \mathcal{D} \left( \hr(t)  \right)
\label{gskl}
\end{equation}
The first term at the RHS in Eq.(\ref{gskl}) is the standard Liouville-Von Neumann quantum commutator describing the coherent, wave-like hopping mediated by the tridiagonal Hamiltonian matrix. The second term, $\mathcal{D}(\hr)$, is the Lindblad Dissipator that tends to destroy quantum coherence.

\subsection{ Matrix Representation of the Electronic Subspace }

The explicit matrix elements in TB representation are written:
\begin{align}
\bra{n} [\hat{H},\hr] \ket{m} &= \bra{n} \hat{H}\hr \ket{m} -  \bra{n} \hr\hat{H} \ket{m} = \nonumber \\
&= \sum_k \left( \bra{n}  \hat{H} \ket{k} \bra{k} \hr \ket{m} - \bra{n} \hr\ket{k} \bra{k} \hat{H} \ket{m} \right) \nonumber \\
&= \sum_k (H_{nk} \rho_{km} - \rho_{nk} H_{km} )
\label{tbeq1}
\end{align}

Isolating the diagonal and off-diagonal terms:
\begin{equation}
\bra{n} [\hat{H},\hr] \ket{m} = (\varepsilon_n - \varepsilon_m) \rho_{nm}  + \sum_{k\ne n} T_{nk} \rho_{km} -  \sum_{k\ne m}\rho_{nk} T_{km} 
\label{tbeq2}
\end{equation}

For $n=m$ the first term is zero, and the evolution depends only on the off-diagonal terms $T_{ij}$ and $\rho_{ij}$. For $n \ne m$ the $(\varepsilon_n - \varepsilon_m)$ term gives a fast oscillation (phase rotation).

\subsection{Structure of the Lindblad dissipator}

The dissipator, explicitly including the two terms "sandwich" and anticommutator, functions as a mathematical "quantum jump" mechanism that captures the interactions with the fluctuating protein environment:
\begin{equation}
\mathcal{D}(\hr) = \sum_{k} \gamma_{k} \left(  \hat{L}_{k} \hr \hat{L}_{k}^{\dag } -  \frac{1}{2} \{ \hat{L}_{k}^{\dag } \hat{L}_{k},\hr \} \right)
\label{dissip}
\end{equation}
 $\hat{L}_{k}$ are the operators representing specific physical mechanisms acting on the protein chain; the $\gamma_{k}$ are relaxation rates that quantify how rapidly a specific dephasing/dissipation channel operates; $\{ \dots \}$ denote the anti-commutator, preserving the total probability trace of the system.

The choice of the Lindblad operators is in principle quite arbitrary, depending on the quantum system to be represented. They must only comply with some physics-induced requirements, such as detailed balance among the up/down channels, charge/number conservation, algebraic closure over the eigenstates of the Hamiltonian, conservation of unitary trace.

The index $k$ in the Lindblad dissipator equation (\ref{dissip}) represents the individual relaxation and dephasing channels operating on the system. We will distinguish lowercase $\gamma_k$ rates for dephasing-type operators, and uppercase 
$\Gamma_{ij}$ rates for relaxation/hopping-type operators between pairs of sites $(i,j)$. However, note that these do not necessarily identify with the simple count of the redox sites. While directly tied to the redox sites (cofactors), the total number of terms in the summation depends on the specific physical mechanisms being modeled. 

For relaxation (that is, the actual movement of the electron), the index $k$ maps to the pathways or transitions between sites, rather than the sites themselves.  
Forward-transfer channels, and matching backward-transfer channels if thermal back-activation is also modeled, must be included. Each term represents an active electron transfer step along the wire. 
Then, to simulate the complete (non-branched) transport chain of length $N$ using a strict Markovian Lindblad approach, the summation over $k$ in equation (\ref{dissip}) should include:
\begin{equation}
\sum _{k} \dots = \underbrace{ \sum_{i=1}^{N} \gamma_{i} \left( \dots \right)}_{\textrm{N dephasing terms}} + 
    \underbrace{ \sum_{\langle i,j \rangle} \Gamma_{ij} \left( \dots \right)}_{\textrm{N-1 forward (+ backward) spatial steps}}
\label{opers}
\end{equation}

Therefore, the index $k$ runs over an ensemble of at least 2$N$-1 to 3$N$-2 distinct operational channels, to capture the complete quantum dynamics of the $N$-site "quantum wire".

\subsection{Lindblad operators for the empirical model}

To describe a chain of tight-binding transport centers with an "empirical" Lindblad model, the interaction of electrons with the protein matrix may be summarized in terms of two primary categories of jump operators:
\vspace{2mm}

\noindent \textbf{i. Pure Dephasing Operators} (phase decoherence).
The protein environment causes the localized site energy levels $\varepsilon_{i}$ to fluctuate rapidly due to thermal motions. This destroys the relative phases between cofactors without exchanging actual energy. The explicit operator is:
\begin{equation}
\hat{L}_{i} = \ketbra{i}{i}
\end{equation}
where $i$ is a specific cofactor state. This operator is self-adjoint, $\hat{L}_i = \hat{L}_i^{\dag}$. Its effect is to selectively wipe out the off-diagonal elements of the density matrix ($\rho_{ij}\rightarrow 0$), dampening the quantum interference(s). It transitions the transport mechanism from a coherent quantum wave, into a classical, incoherent hopping model.

The so-called "sandwich" matrix element is:
\begin{equation}
\bra{n} ( \hat{L}_{i} \hr \hat{L}_{i}^{\dag}) \ket{m} = \bra{n} ( \ket{i} \bra{i} \hr \ket{i} \bra{i} ) \ket{m} 
\end{equation}
and by orthonormality of the basis functions:
\begin{equation}
\bra{n} ( \hat{L}_{i} \hr \hat{L}_{i}^{\dag}) \ket{m} = \rho_{ii} \, \delta_{ni} \, \delta_{mi} 
\end{equation}
This is zero everywhere but in the single diagonal point where the electron is on the fluctuating site $\ket{i}$.

For the "decay" anticommutator, since the product $\hat{L}_i^{\dag}\hat{L}_i = \ket{i}\braket{i}{i} \bra{i} = \ketbra{i}{i}$, we have:
\begin{align}
-\frac{1}{2} \bra{n} [\hat{L}^{\dag}_i\hat{L}_i, \hr] \ket{m} &=
-\frac{1}{2} \bra{n} ( \ketbra{i}{i} \hr + \hr \ketbra{i}{i} ) \ket{m} = \nonumber \\
&= -\frac{1}{2} (\delta_{ni} \rho_{im} + \rho_{ni} \delta_{im})
\end{align}

\noindent \textbf{ii. Relaxation Operators } 
(Energy Dissipation). These terms account for the irreversible downhill physical transport of the electron down the redox gradient toward the terminal oxidation cluster, dropping its excess energy in the form of vibrational heat (phonons) into the protein. The explicit operator is:
\begin{equation}
\hat{L}_{j \leftarrow i} = \ketbra{j}{i}
\label{1way}
\end{equation}
representing an electron jumping from site $i$ to site $j$. 

The effect of such operators is to directly modify the diagonal populations $\rho _{ii}$ of the density matrix, reflecting the true directional current across the chain. Matrix elements are calculated, for the "sandwich" term:
\begin{align}
\bra{n} ( \hat{L}_{j \leftarrow i} \hr \hat{L}_{j \leftarrow i}^{\dag}) \ket{m} &= 
\bra{n} ( \hat{L}_{j \leftarrow i} \hr \hat{L}_{i \leftarrow j}) \ket{m}  = \nonumber \\
&= \bra{n} ( \ket{j} \bra{i} \hr \ket{i} \bra{j} ) \ket{m} = \nonumber \\
&= \rho_{ii} \, \delta_{ni} \, \delta_{mj} 
\end{align}
If before the jump the electron was on the starting site $i$ with $\rho _{ii}$, this term is a source that injects electron population in the landing site  $j$, changing the diagonal element $\rho _{jj}$.

For the "decay" anticommutator, the product $\hat{L}^{\dag}_{j \leftarrow i}\hat{L}_{j \leftarrow i} = \ket{i}\braket{j}{j} \bra{i} = \ketbra{i}{i}$, and we have:
\begin{align}
& -\frac{1}{2} \bra{n} [\hat{L}^{\dag}_{j \leftarrow i}\hat{L}_{j \leftarrow i}, \hr] \ket{m} = \nonumber \\
= & -\frac{1}{2} \bra{n} ( \ketbra{i}{i} \hr + \hr \ketbra{i}{i} ) \ket{m} = \nonumber \\
= & -\frac{1}{2} (\delta_{ni} \rho_{im} + \rho_{ni} \delta_{im})
\end{align}

\noindent \textbf{iii. Sink Operators} Finally, we must add a "sink" operator $\hat{L}_S$ at the terminal cluster, to enforce the biophysical condition of final electron transfer out of the complex. Two alternative choices are used in the literature to this effect. One can either add an explicit extra site (such as the quinone in Complex-I) with a sink constant $\gamma^{\text{sink}}$ typically a few meV,
\begin{equation} 
\hat{L}_S = \sqrt{\gamma^{\text{sink}}} \, \ket{N+1}\bra{N}
\end{equation}
When the electron hits site $N$ it is "sucked" to site $N+1$: since there is no inverse operator $\vert N \rangle \langle N+1 \vert$, the electron can no longer re-enter the chain.

The alternate choice is to use a non-Hermitian absorption operator on the terminal site,
\begin{equation} 
\hat{L}_S = -i \, \hbar \, \gamma^{\text{sink}} \, \ket{N}\bra{N}
\end{equation}

In practice, the two solutions give numerically identical results except at the last site, and the choice is only a matter of convenience. In the non-Hermitian case, the $N$-site density matrix does not conserve the trace by construction (because of the last site). Therefore, for heat and entropy calculations  the explicit-sink version may be a cleaner thermodynamic quantity to use.
\vspace{2mm}

The advantage of such a Lindbladian "memoryless" quantum master equation formulation is that it naturally preserves complete positivity. Alternative methods (like the Redfield or Caldeira-Leggett equations) may produce unphysical, negative electron populations when applied to complex "biological" topologies. The Lindblad structure guarantees that populations stay mathematically valid between 0 and 1 throughout the entire simulation. However, as already said, its domain of applicability is restricted to conditions of weak electron-bath coupling. The numerical implementation of this "empirical" Lindblad model, termed EL in the following, is described in Appendix \ref{workf}.

\section{Lindblad equations with thermodynamic consistency}\label{lindterm}

The Lindblad equation, while a fundamental tool for open quantum systems, is not well tailored to work at finite temperatures.
The positivity of the density matrix $\rho$ is not sufficient to guarantee thermodynamic consistency. Positivity only describes the statistical consistency of quantum states, but it completely ignores energetic and entropic constraints. A density matrix can be perfectly positive while describing a system that absorbs heat from a reservoir and converts it into work. In fact, when deriving the Lindblad equation in the Born-Markov approximation, the jump operators are calculated based on the instantaneous (memoryless) transition frequencies of the isolated system. This guarantees relaxation to a steady state, but does not guarantee that this state is the correct Gibbs state, unless some form of the quantum detailed balance condition is respected. For this reason, additional conditions must be imposed to the operators to also satisfy thermodynamics consistency (the Gibbs distribution and the second law), either by construction, or by supplementary constraints on the dissipation/relaxation operators.

\subsection{Dissipation operators from the quantized Klein-Kramers equation}

In contrast to the standard "empirical" approach with phenomenological dissipation, we recently established a more rigorous foundation for open quantum systems \cite{giord0,giordan} (called "BtL"), which allows to systematically deriving Lindblad-type master equations by applying canonical quantization to the classical Klein-Kramers framework. By incorporating friction and noise symmetrically into the classical Hamilton equations for both position and momentum, we could generate completely positive, trace-preserving quantum dynamics. It was shown that such a microscopic derivation ensures full thermodynamic consistency, satisfying both the first and second laws of thermodynamics via the monotonicity of quantum relative entropy, and provides a robust, universally applicable toolkit for modeling out-of-equilibrium nanoscale systems without risking unphysical negative probability states.

Using the results of the BtL paper, we will plug exactly that formalism in the above Lindblad structure. In  \cite{giordan}, the formalism was explicitly written for the case of a quantum harmonic oscillator. The same formal development can be translated into a site-localized model, however with the important note that in this case the $\hq$ and $\hp$ operators are no longer canonical variables of the electronic Hamiltonian. The site index $i$ is not be interpreted as an oscillator quantum number, because sites are different, spatially separated redox centers: we changed from a representation of one site with $N$ energy levels, to a representation of $N$ sites with one energy level each. The two operators rather describe the fluctuations of a continuous charge distributed over the entire chain. By way of the bosonic operators $\cc,\cd$, a fluctuation of density $\rho_k = \sum_i \cd_{i+k}\cc_i$ displaces charge along the chain. In this sense, such a model is a kind of continuum limit of the discrete electrons described by the individual jump operators, rather describing a mean-field dynamics. 

As shown in Appendix \ref{giordx}, Lindblad operators can be conveniently cast in the form:
\begin{align}
 \hat{L}_1 &= a \, \hq \,\, + \,\, i \,b \hp = \sqrt{\lambda_1} \, (\cos\theta \, \hq + i \, \sin\theta \, \hp)  \nonumber \\
 \hat{L}_2 &= c \, \hq \,\, + \,\, i \, d \hp = \sqrt{\lambda_2} \, (-\sin\theta \, \hq + i \, \cos\theta \, \hp) 
\label{gem2}
\end{align}
where the explicit coefficients $a,b,c,d$ are obtained from the eigenvalues of Kossakowski's matrix construction, Eq.(\ref{kossa}),
\begin{equation}
\lambda_i = - (A+C) \pm 2 \sqrt{(A+C)^2 + (B-D)^2 / \hbar^2 } 
\end{equation}
with $\pm$ for $i$=1,2, $\theta = \tfrac{1}{2} \arctan [ (B-D) / \hbar(C-A) ]$,
and  $A,B,C,D$ given by Eq.(\ref{ABCDdef}).

As said, in this BtL formulation, friction and noise act symmetrically on both Hamilton equations for the position coordinates $q$ and the momenta $p$. In a redox chain of tunneling sites, this means that one cannot simply model the electron jump just as a population decrease (acting only on the space of localized states, equivalent to the momentum). Dissipative couplings aree included for both the spatial decomposition operators (related to the coordinate of the cofactors and the potential gradients), and the momentum/energy decomposition operators. 
Thermal detailed balance is not imposed empirically, but is included by construction in the structure of friction operators with the hyperbolic 
function $\Theta(\omega,T)$, Eq.(\ref{giord}). 

The main advantage of such a "continuum charge" model will be to expose the characteristic  physics-thermodynamics limits of the "empirical" Lindblad approach (EL). However, the corresponding description of charge dynamics may be too much constrained by the construction.

\subsection{Empirical Lindblad model with quantum detailed balance}\label{lindtherm}

Comparison of the EL and BtL models (see Results section \ref{risulta}) will show that including thermal detailed balance in quantum dynamics is essential to preserve consistency with the basic tenets of thermodynamics, which regulate heat and entropy exchanges with the external bath. 

As implied by Eq.(\ref{opers}), the operator structure of the empirical model can in principle include both forward and backward jump events (as well as other kinds). In a thermodynamically safer, yet still phenomenological model, we can impose local detailed balance on thermal jump rates. For every $i,j$ pair of adjacent sites, introduce:
\begin{align}
\hat{L}_{j \leftarrow i} &= \sqrt{k_{ji}} \, \ketbra{j}{i} \nonumber \\
\hat{L}_{j \rightarrow i} &= \sqrt{k_{ij}} \, \ketbra{i}{j}
\end{align}
with the condition:
\begin{equation}
\frac{k_{ij}}{k_{ji}} = e^{-\frac{(\varepsilon_j - \varepsilon_i)}{k_BT}}
\label{thermratio}
\end{equation}
with the $k_{ij}$ the temperature-renormalized off-diagonal terms corresponding to the $T_{ij}$ of the bare electron.

These operators replace the one-way relaxation operators Eq.(\ref{1way}) in the empirical Lindblad model, while the dephasing and sink operators remain the same. Now, this choice ensures that the only possible stationary state for the dissipator is the Gibbs thermal state.

This may appear as an entirely "ad hoc" choice, to make the jump rates temperature dependent. However, it has a deeper justification grounded in the polaron model \cite{krause15,rouse22,manzano}. In fact, starting from Eq.(\ref{hamil}), the phonon "cloud" can be used to dress the jump operators:
\begin{equation}
\ketbra{i}{j} + \ketbra{j}{i}  \longrightarrow \ketbra{i}{j}  \otimes B_{+} + \ketbra{j}{i} \otimes B_{-}
\end{equation}
with the "phonon cloud" displacement/rearrangement operators
\begin{equation}
B_\pm = \exp\left[ \pm \sum_k \frac{g_k}{\omega_k} (b_k^\dagger - b_k) \right]
\end{equation}
meaning that the electron jump is correlated to the protein displacement or structural rearrangement, $\hat{b}_k^{\dag}, \hat{b}_k^{ }$, with el-ph coupling constants $g_k$ and vibronic frequencies $\omega_k$.

The states of the density matrix are no longer reduced to the naked electronic $\ket{i}$, but are true polaronic states $\ket{i,\mathbf{0}_i}$ where $\mathbf{0}_i$ are phonon baths relaxed in different geometrical configurations.
The electronic site energies are screened by the Marcus'  reorganization energy, $\bar{\epsilon}_i = \epsilon_i - \lambda$, with
\begin{equation}
\lambda =\sum_ k \frac {|g_k|^2}{\omega_k}
\end{equation}
The new jump rates are time integrals of the correlation function of the protein displacement/rearrangement operators 
\begin{equation}
\Gamma _+ = k_{ij} \int _{-\infty }^{\infty } dt \, e^{i( \bar{\epsilon}_i - \bar{\epsilon}_j) t } \langle B_{-}(t) B_{+}(0) \rangle_B
\end{equation}
and the $k_{ij} \sim T_{ij} \exp(-\lambda/2k_BT)$.

By developing this expression at biological temperatures ($k_B T \gg \hbar\omega$), it is shown that the integral ends up in the classical Marcus' equation for electron transfer \cite{krause15}:
\begin{equation}
\Gamma _+ \approx \frac{k_{ij}}{\hbar} \sqrt{\frac{\pi}{\lambda k_{B}T}} \exp \left( -\frac{(\Delta \bar{\epsilon}+\lambda)^2}{4\lambda k_BT} \right)
\end{equation}
 
Now, the forward/backward jump ratio of the Marcus' terms is:
\begin{align}
\frac{\Gamma_+}{\Gamma_{-}} &= \exp \left[ -\frac{1}{4\lambda k_BT} 
\left( (\Delta \bar{\epsilon} + \lambda)^2 - (-\Delta \bar{\epsilon} + \lambda)^2 \right)
\right]  \nonumber \\
&= \exp \left( -\frac{4\Delta \bar{\epsilon} \lambda}{4\lambda k_BT} \right) = \exp \left( -\frac{\Delta \bar{\epsilon}}{k_BT} \right)
\end{align}
that is, Eq.(\ref{thermratio}) above with $\Delta\bar{\epsilon} = (\epsilon_i - \lambda) - (\epsilon_j - \lambda) = \epsilon_i - \epsilon_j$.
Therefore, in this formulation the reorganization energy $\lambda$ no longer does appear explicitly.

This thermally-adjusted "polaron" model, which will be labelled TEL (thermal-empirical Lindblad) in the foregoing, combines coherent dynamics, from the Hamiltonian off-diagonal elements that maintain quantum coherences (polaron delocalization) at short times, and thermal relaxation. The Lindblad heat sink, thanks to the Boltzmann ratio, will drive the system towards the correct thermal equilibrium state, destroying superfluous coherences at long times without violating the second law of thermodynamics.

\section{Results}\label{risulta}

Since the different models do not share common parameters to impose a common time scale, we can compare them for example by adjusting the time plots of the respective extraction rate Q ($=\hr_{99}$, "9" being the $N$+1-th sink site), trying to match two reference values $\hr_{99}\simeq 0.2$ at $t$=5 and $\hr_{99}\simeq 0.8$ at $t$=10 (in arbitrary units, see note in Appendix \ref{workf}). For further comparison, a Pauli-like, classical master equation model can also be obtained from the EL model, by setting all the tunneling parameters $T_{ij}=0$ and suppressing the off-diagonal elements of the dissipator $\mathcal{D}$. In all models, the initial site population is $\hr_{11}=1$ and $\hr_{ij}=0$ for $i,j\neq 1$ at $t$=0.  

\subsection{Site populations from the density matrix}

\begin{figure}[t]
\centering
\includegraphics[width=\columnwidth]{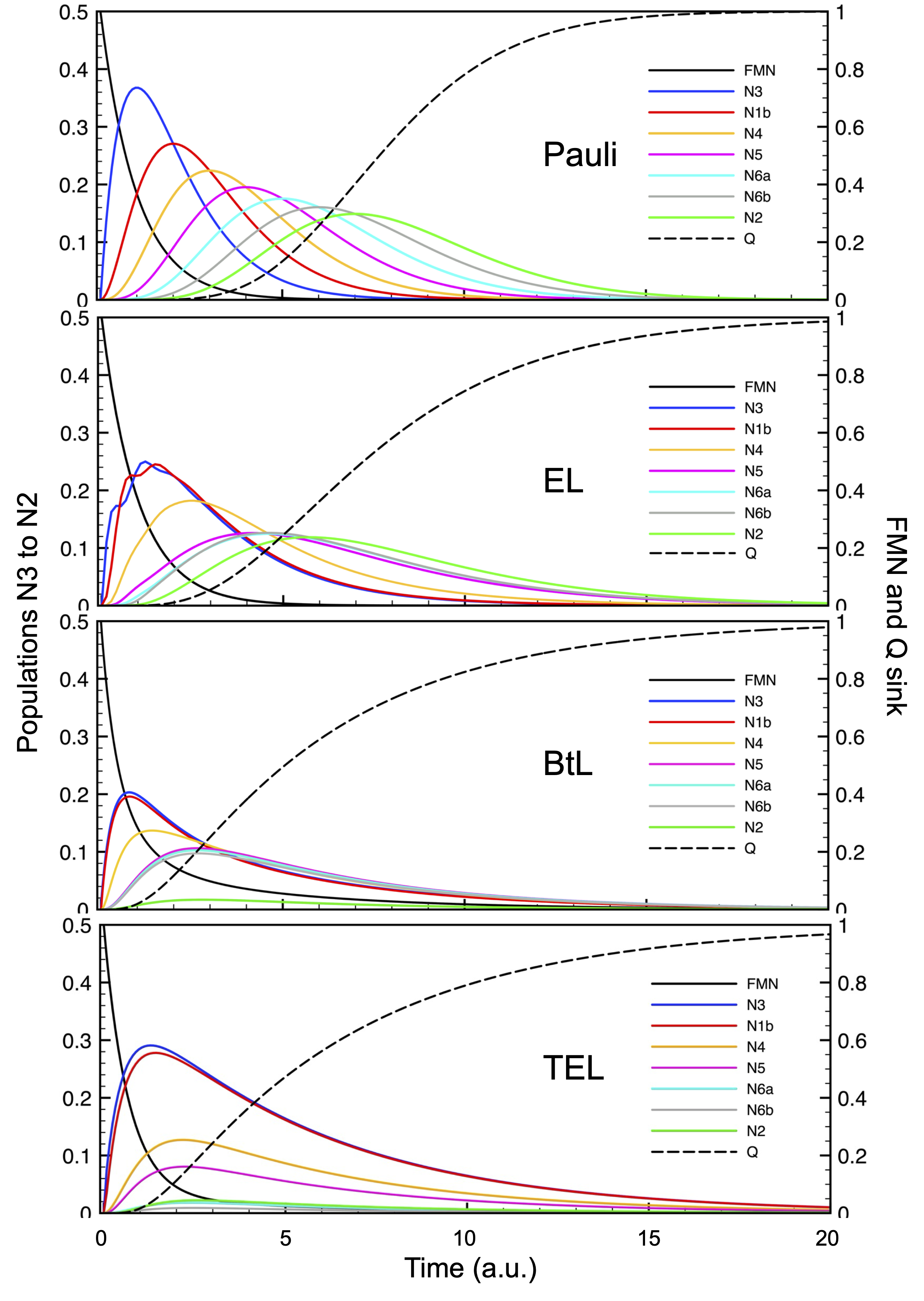}
\caption{Comparison of the 8-site (mock-up model of Complex I) population behavior, between the classical master equation (Pauli), and the quantum empirical Lindblad (EL), BtL and thermal-empirical Lindblad (TEL) models. Plots approximately normalized at Q20 and Q80 (see text). Note that the input / output channels FMT and Q (black full/dashed line) are plotted to the right y-axis.}
\label{fig1}
\end{figure}

The local occupation of each redox site can be calculated explicitly from the diagonal terms $\hr_{ii}(t)$ of the density matrix. Note that for the EL model this quantity is truly a single-particle property, whereas for the BtL model it rather represents the flux density of the transported charge. The first example is an 8-site model, with values of $\varepsilon_i, T_{ij}$ roughly approximating those of mitochondrial Complex I. Figure \ref{fig1} shows comparable time evolutions for the Pauli, EL, BtL and TEL models, where a quite different dynamics can be immediately observed.

The classical master equation (top panel) describes electrons as particles physically "hopping" over distant sites, with distinct peaks that follow each other in time. Some degree of superposition can be expected also classically, for example with identical transition rates the population of site $n$ should have an Erlang/Gamma-like shape $p_n(t) \sim (k t)^n \exp(-k t) / n!$, with a maximum near  $t \sim n/k$ and spread width of order $\sqrt(n)/k$. So, neighboring populations can overlap, even quite strongly for small $n$. 

In the quantum EL model (second panel), the probability densities are overlapping, with very asymmetric rise/decay times, peaks broadened or sharpened by the different rates. The density matrix evolves as a superposition/mix of states over the whole chain (probability + coherent states). The Hamiltonian couplings spread amplitudes coherently over several sites, while the dissipator dephases/relaxes it, so at a given time the population is distributed over multiple cofactors.

The EL model is mostly unidirectional, so it is more like an effective driven transport model than an equilibrium thermal bath. On the one hand, such a representation is useful because it is kinetically expressive, it describes single particles crossing sites, all the 
$\gamma$ constants can be separately tuned. On the other hand, thermodynamically it does not represent a bath at temperature $T$. A generic set of jump/dephasing rates can be completely positive and trace-preserving while still violating detailed balance, entropy-production positivity, or heat-flow consistency.
Also, notice the small wiggles at short times in $\rho_{22},\rho_{33}$, to be further discussed in section \ref{current}.

Compared to the EL, the thermally-reversible BtL (third-lower panel) appears constrained,  because the sites are not just fed and emptied sequentially (by the $\hat{L}_{j\leftarrow i}$ relaxation operators), but are locally equilibrating with neighbors. This makes the $\rho_{ii}$ curves more nested, each one appearing enveloped under the preceding one. Such different behaviors of the quantum models are very robust and do not change drastically by changing the scaling parameters $m,\omega,\beta$.
In the BtL model two-way quantum coherence comes into play, by coupling the electron to the protein bath through linear combinations (a kind of bath-induced quantum viscosity).
As the electronic wave advances along the quantum wire, quantum dissipation $Q(\omega,T)$ and dephasing $k_{B}T$ act in parallel, constantly dispersing the phase information. The wave undergoes progressive spatial damping: the probability is progressively distributed among the sites, rather than being transferred in discrete lumps. This is a key difference: the EL model describes Complex I as a ladder where the electron is a falling from one rung to the next in nearly-discrete steps. The BtL model, instead, describes Complex I as a "fluid" quantum channel, where the electron is a wave packet that delocalizes on the central plateau and slides smoothly toward the exit, geometrically damped by the friction of the protein. 

Interestingly, the TEL "polaron"-like model (lowest panel) gives results that look more close to those of BtL than to the empirical EL. Also in this case we see all the occupation plots for the chain sites to be enveloped into one global behavior, with maxima only slightly displaced in time. In this case, it is rather the integrals of each curve that tell a different story, by decaying exponentially. These integrals are proportional to the dwelling time of the electron, $\tau_i \propto \rho_{ii}(t) dt$, and 
are of order $\tau_i \sim (\sqrt{k_{ij}} / \Delta \bar{\epsilon})^{2i}$, whence their exponential decay along the chain.

\begin{figure}[t]
\centering
\includegraphics[width=0.85\columnwidth]{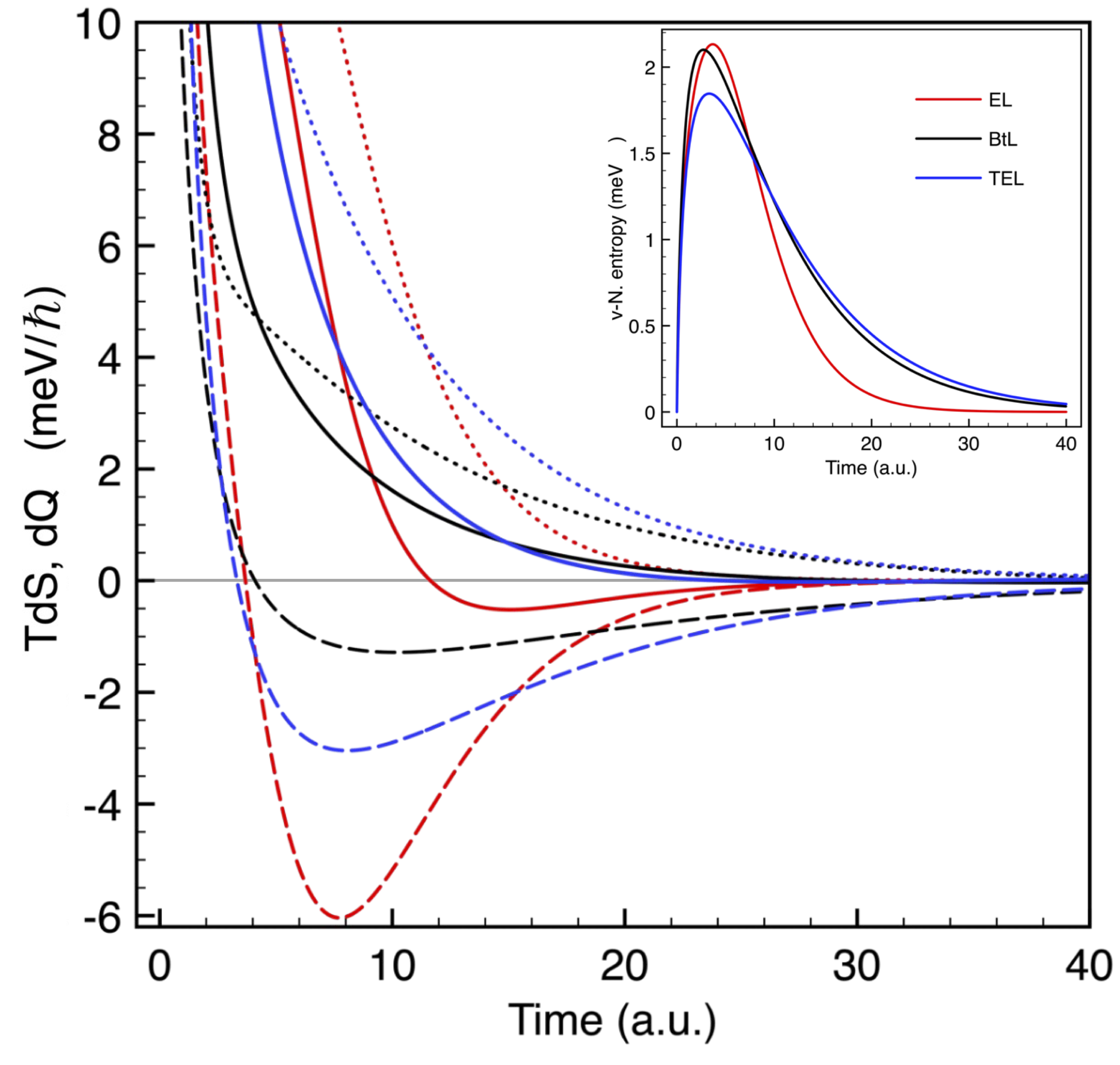}
\caption{Plot of the thermodynamic quantities for similar values of the charge flow dynamics. Red plots = EL model, black plots = BtL model, blue plots = TEL model. Full lines, check of the second principle, $TdS - dQ$; dotted lines, heat released to the environment, $-dQ$; dashed lines, von Neumann entropy multiplied by $k_BT$. The inset verifies the positivity of the entropy for the three models. (Same numerical parameters as in Fig. \ref{fig1}.)}
\label{fig2}
\end{figure}

In the semi-classical (Pauli) or weakly-coupled EL models, the redox chain is seen as a series of classical "boxes" in which the charge moves hydraulically: the electron accumulates at site 1, then empties almost completely at site 2 (generating a sharp maximum), then at site 3, and so on. This happens because these models assume that the intermediate sites can "hold" the bare electron for an appreciable time before the next jump occurs.
In the TEL model, instead, the charge density is delocalizing coherently under the effect of the initial polaron transformation, but the interaction with the polaronic bath causes the probability of measuring the electron stationary at the advanced sites to decay exponentially with the distance from the initial donor. All the curves fall under the same envelope because the dynamics is no longer a macroscopic sequential cascade, but the decay of an initial polaronic state that projects its quantum tail to the subsequent sites.

\subsection{Thermal dynamics}

To compute the instantaneous heat flux $\frac{dQ}{dt}$ that the electron transfers to the protein, we apply the First Principle in the quantum version for open systems. The average energy of the electron is the expectation value of the electron Hamiltonian, $E(t) = \text{Tr}(\hat{\mathcal{H}}_0 \hat{\rho}(t)$. This must be derived with respect to time:
\begin{equation*}
\frac{dE(t)}{dt} = \textrm{Tr} \big[ \hat{\mathcal{H}}_0 \frac{d\hr}{dt} \big]
\end{equation*}
But replacing the $d\hr/dt$ by its Lindblad eq. expression, the first term with the commutator is $\textrm{Tr} ( \hat{\mathcal{H}}_0, [\hat{\mathcal{H}}_0, \rho] ) = 0$, and the heat is only given by the dissipator term 
\begin{equation}
\frac{dQ(t)}{dt} = \textrm{Tr} \big[ \hat{\mathcal{H}}_0 \mathcal{D}(\hr) \big]
\label{energio}
\end{equation}
Our sign convention is that for $\frac{dQ}{dt} < 0$ the electron is losing (potential or kinetic) energy and transfers it to the protein, and the opposite occurs for $\frac{dQ}{dt} > 0$ (endothermic process).

By the Second Principle, we can calculate also the heat transfer via the entropy production induced by the interaction with the heat bath. Formally, we can apply the time derivative to the von Neumann entropy 
\begin{equation}
\frac{dS}{dt} = - \frac{d}{dt} \text{Tr}(\hr \ln \hr) = - \text{Tr} \left( \frac{d\hr}{dt} \ln \hr \right)
\end{equation}
(by using the conservation of trace, $\text{Tr}(\hr)=1$).

Figure \ref{fig2} shows the $TdS-dQ$ for the three models EL, BtL and TEL (red, black and blue continuous lines, respectively), together with the separate components ($TdS$ long-dashed, $dQ$ dotted lines). It can be seen that the EL model gives a $TdS<dQ$ over a large time interval (full red line), demonstrating the lack of thermodynamic consistency. Both the BtL and the TEL models properly converge to thermal equilibrium without apparent violations up to long times (within the numerical integration error). The total entropies, shown in the inset, remain positive, however the faster decay of the EL entropy (red) betweeen $t$=10-20 is likely responsible for the violation of the second principle. On the other hand, it is seen that the TEL entropy closely follows the BtL profile, apart from a slightly lower maximum value. 

Overall the BtL model, although not stemming from a fully realistic depiction of the electron wavefunction, sets quite rigorous bounds on the thermal consistency of the coupled system's evolution. The TEL model appears to be able to implicitly recover these bounds, turning the empirical Lindblad approach into a more consistent picture of the electron transport, when considering typical values of the on-site and tunneling energies of the Complex I system. This will be even more strongly confirmed by looking at the effects of quantum coherence in the next sections.

\begin{figure}[t]
\centering
\includegraphics[width=\columnwidth]{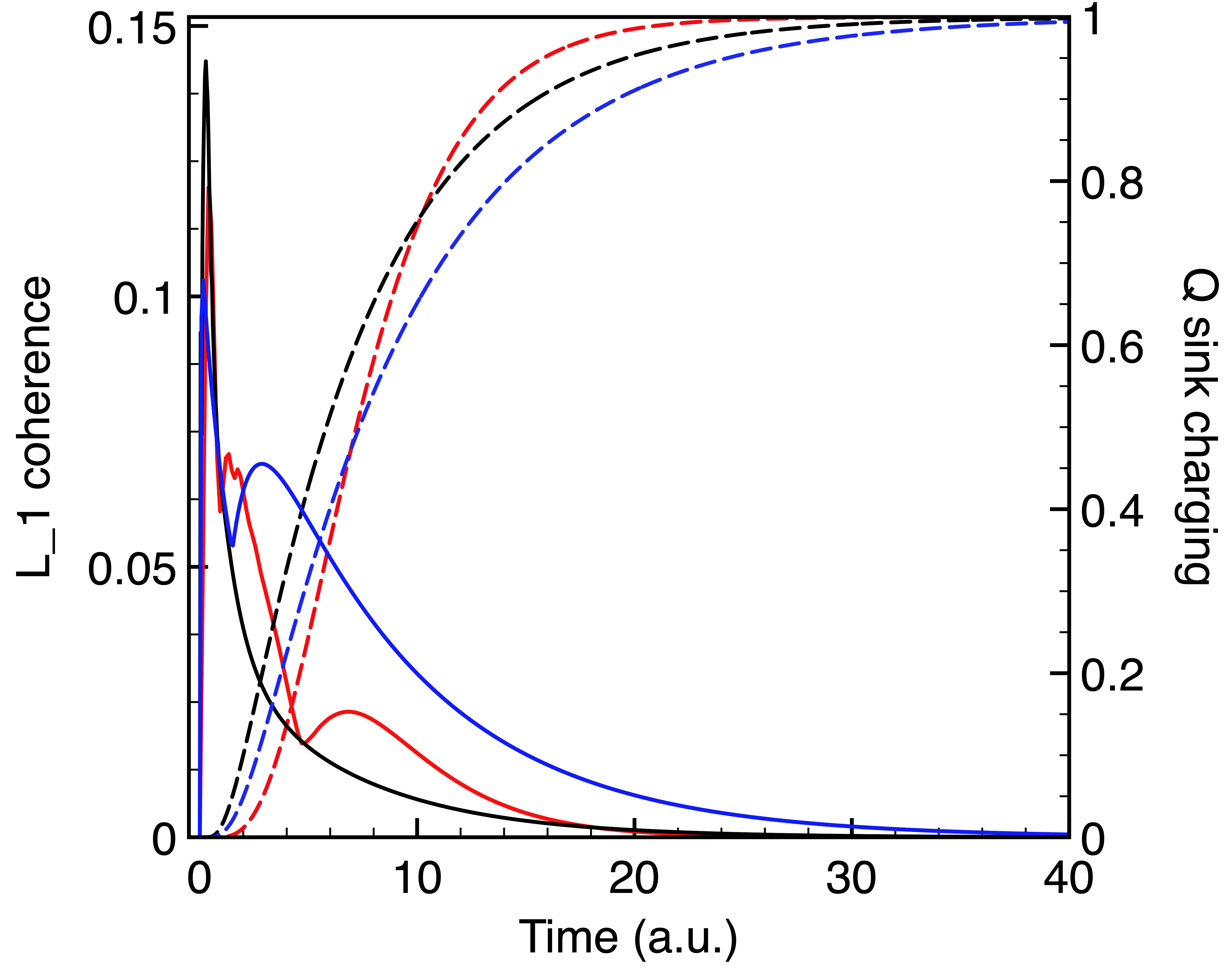}
\caption{Plot of the quantum coherence (full lines, left y-axis) and Q sink charging (dashed lines, right y-axis). Red plots = EL model, black plots = BtL model, blue plots = TEL model.}
\label{fig3}
\end{figure}

\subsection{Quantum coherence}

To quantify the total coherence of the system at any instant, the most widely used metric both in information and quantum biology is the $l_1$ norm of coherences, $\mathcal{C}_{l_1}$. This measure is simply calculated by adding the moduli of all off-diagonal elements of the density matrix:
\begin{equation}
\mathcal{C}_{l_{1}}(t) = \sum _{i\ne j} |\rho _{ij}(t)|
\end{equation}

The theoretical maximum value of quantum coherence is easily calculated, for an $N$ site system describing the transit of a single electron at a time has a real dimension of $N$ (it would be $2^N$ in a complete Fock space that can be occupied by more than 1 electron).
Our system is maximally coherent (pure delocalized state) when the electron is "spread" equally over all the sites at the same time with the same phase, $\ket{\psi _{max}} = \tfrac{1}{\sqrt{N}} \sum_i \ket{i}$.  
In this asymptotic scenario, each element of the density matrix is $\rho_{ij} = 1/N$. For the $N$=8 case, the maximum value of norm $\mathcal{C}_{l_1}$ for the 64-8=56 off-diagonal elements is therefore 56$\cdot$0.125 = 7. Lower values will indicate a reduced correlation, progressively dispersing over time.

Figure \ref{fig3} shows the evolution of quantum coherence for the EL, BtL and TEL models, together with the terminal quinone sink filling (dashed lines). The comparison is useful to estimate the persistence of quantum correlations along the progressive transfer of the electron from the injection point to the terminal site of the chain. The maximum values at the initial peak are $\sim 0.15$ for all the models, that is a rather small overall coherence relative to the maximum value. However, the TEL model displays the longer persistence, with about half of the initial coherence still surviving at 50\% filling (that is, about half of the electron total transit time).

\begin{figure*}[t]
\centering
\includegraphics[width=\textwidth]{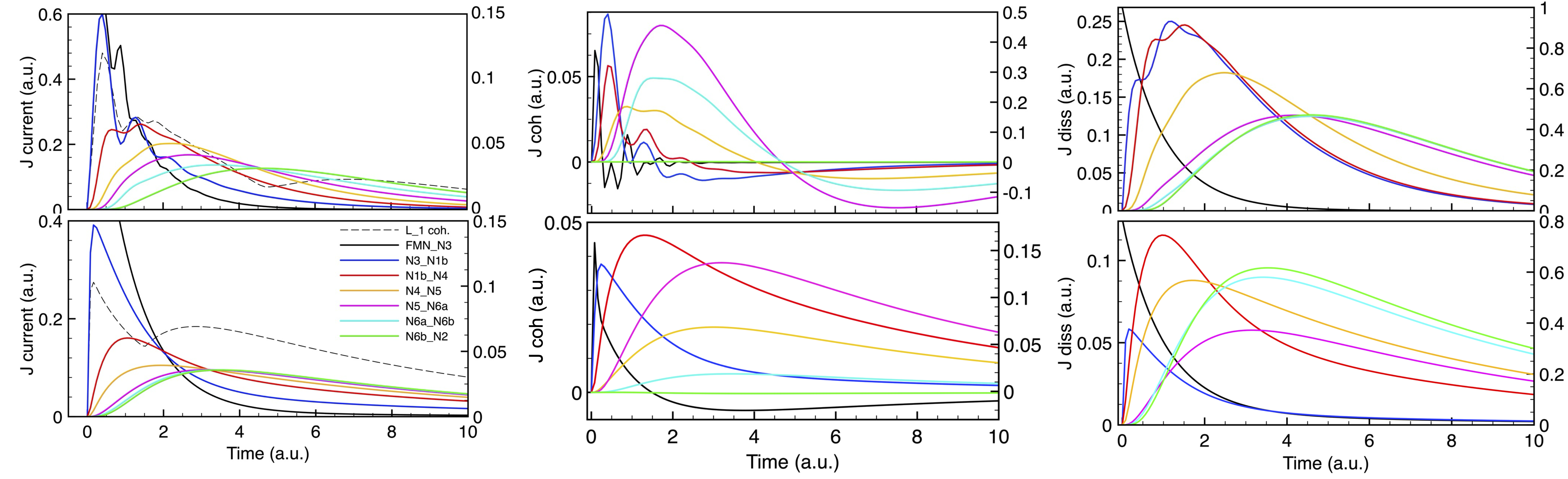}
\caption{Comparison of the $J_{ij}$ currents for the EL (top rows) vs the TEL model (bottom rows). The left panels show the total currents, with color codes for each site in the lower figure; for the sake of comparison, the dashed lines also report the $l_1$ coherences from Fig.\ref{fig3}, to be read on the right y-axis. The central and right panels respectively show the $J^{coh}$ and $J^{diss}$ components, for the EL (above) and TEL (below) models; for the central and right panels, the black and blue plots for $J_{12},J_{23}$ correspond to the right-side y-axis.}
\label{fig4}
\end{figure*}

\subsection{Site currents from the quantum model}\label{current}

The total current is the time derivative of the number operator $\hat{n}$ between two sites
\begin{align}
\hat{J} &= \frac{d \langle \hat{n} \rangle}{dt} = \textrm{Tr} \left[ \hat{n} \left( -\frac{i}{\hbar} [\hat{H},\hat{\rho}] + \mathcal{D}[\hat{\rho}] \right) \right] = \nonumber \\
&= \frac{i}{\hbar} \textrm{Tr} \left( [\hat{H},\hat{n}] \hat{\rho} \right) + \left( \hat{n}] \mathcal{D}[\hat{\rho}] \right) =
 \hat{J}^{coh} + \hat{J}^{diss}
\end{align}

The two contributions are the coherent term from the Hamiltonian phases between $i$ and $j$:
\begin{align}
J_{ij}^{coh}(t) &= \frac{ie}{\hbar} [ T_{ij} \rho_{ij}(t) - T_{ji} \rho_{ji}^*(t) ] = \nonumber \\
&= -\frac{2e T_{ij}}{\hbar} \textrm{Im} [ \rho_{ij}(t) ]
\end{align}
and the incoherent term from the Lindblad dissipator:
\begin{equation}
J_{ij}^{diss}(t) = -e \Gamma_{(d,r)} [ \rho_{ii}(t) - \rho_{jj}(t) ]
\end{equation}
where $\Gamma$ generically indicates the appropriate dissipation/relaxation $(d,r)$ coefficient for the different models.

Figure \ref{fig4} shows the evolution of the site-site $(i,i+1)$ currents for the EL vs. TEL model. The left panels display the total current (zoomed over the time interval [0-10]). Apparently, both EL and TEL give a similar picture, describing charge flowing in smooth packets with separated maxima between the neighboring centers, in particular, the plot for the EL charge distributions closely maps the site population distribution already shown in Fig.\ref{fig1}. However, when looking at the separate components we see a different story. 

The central and right panels respectively display the Hamiltonian coherent contribution and the dissipative contributions to the total current. The TEL model (lower panels) shows quantum coherence developing and decaying in a physically understandable manner: the coherent part grows rapidly in the sites closer to the source, and is canceled by the dissipative part in the sites closer to the terminal sink, in agreement with the decay of overall coherence already seen in Fig. \ref{fig3}. The two contributions are comparable in amplitude, $\sim$0.05-0.1 in the reduced units. Only the first "charge injection" site (representing FMN, in the analogy with Complex I) shows a negative undershoot of the coherent current (see black line, central-lower panel), because of the artificial initial boundary condition that forces the quantum phase without any reaction from the rest of the system downstream. 

On the other hand, the EL model (upper panels) shows a quite unphysical behavior of the current. The coherent part shows wild positive-negative oscillations at all centers, that are suppressed in the total current only by the corresponding much larger dissipative terms. These already showed up also as wiggles at short times, in the population plots of Fig.\ref{fig1}.
Such an asymmetric interaction corresponds to strong coupling with an artificial environment. Mathematically, EL attempts to preserve the positivity of the density matrix, but in doing so it generates artificially strong quantum coherences in the 
$\rho _{ij}$, and negative coherent oscillations or reflections. It is a formal paradox: to force classical unidirectionality in a weakly coupled channel, the model activates artificial quantum resonances that do not exist in the actual biological system, rendering the picture of discrete quantum packets advancing by hopping far less realistic.

In the empirical EL model, the Lindblad terms enforce directed population transfer, giving the impression of more important quantum effects (despite the lower persistence of coherence, Fig.\ref{fig3}), but do not necessarily encode a corresponding thermal decoherence mechanism. As a result, residual coherent recurrences in the plot of $\mathcal{C}_{l_1}$ can appear even at late times, when the population dynamics is already dominated by irreversible extraction. These late coherence bumps should therefore be interpreted as model-induced coherences (or artifacts), rather than direct evidence for persistent quantum transport.

\section{Some observable consequences of quantum effects in biophysical redox chains}

In this final section, we use the TEL model to show two theoretical effects that could turn into physical consequences of quantum behavior, possibly observable in such tightly coupled tunneling-hopping electron transport systems; namely, the role of bottlenecks in on-site energies, and the possibility of super-exchange mechanisms to bypass them.

We use again a $N$=8 site toy model of the electron transport in the "polaron"-like TEL model, which based on the results of the previous sections, appears to give the most complete description of the coupled quantum-dissipative physics. We start with flat values of all the onsite energies, except the first and last site, $-\epsilon_i = [0,u,u,u,u,u,u,2u]$, and flat values for the tunneling energies, $T_{i,i\pm1} = u/20$. Also, the dephasing/relaxation and sink constants are fixed to equal values, $\gamma_i=1$, unless otherwise noted. Such relative scales of values are typical of biophysical redox chains, for example in mitochondrial Complex I, for which $u\simeq$80-90 meV. Since $e=\hbar=1$ in our codes, time units are arbitrary.

\subsection{Quantum bottlenecks}

Already in the early works examining the on-site energies of Complex I from the donor (NADH) to the acceptor (quinone) \cite{stuche1,stuche2}, it had been observed that the energy profile is not a smooth, linear descent. Rather, the chain exhibits a "rollercoaster" profile, in which certain intermediate clusters possess energies significantly higher (barriers) or lower (traps) than those of their neighbors. Such alternating energy barriers are often found in biophysical redox chains \cite{guidoni}. In particular, Stuchebrukhov identified the transition between clusters N5 and N6a in Complex I as having the largest geometric separation (approximately 14 \AA) and an unfavorable energy gap. His initial calculations indicated that pure quantum tunneling would take far too long to bridge the gap, compared to experimental macroscopic data ($\approx 90-200 \mu$s).

\begin{figure*}[t]
\centering
\includegraphics[width=\textwidth]{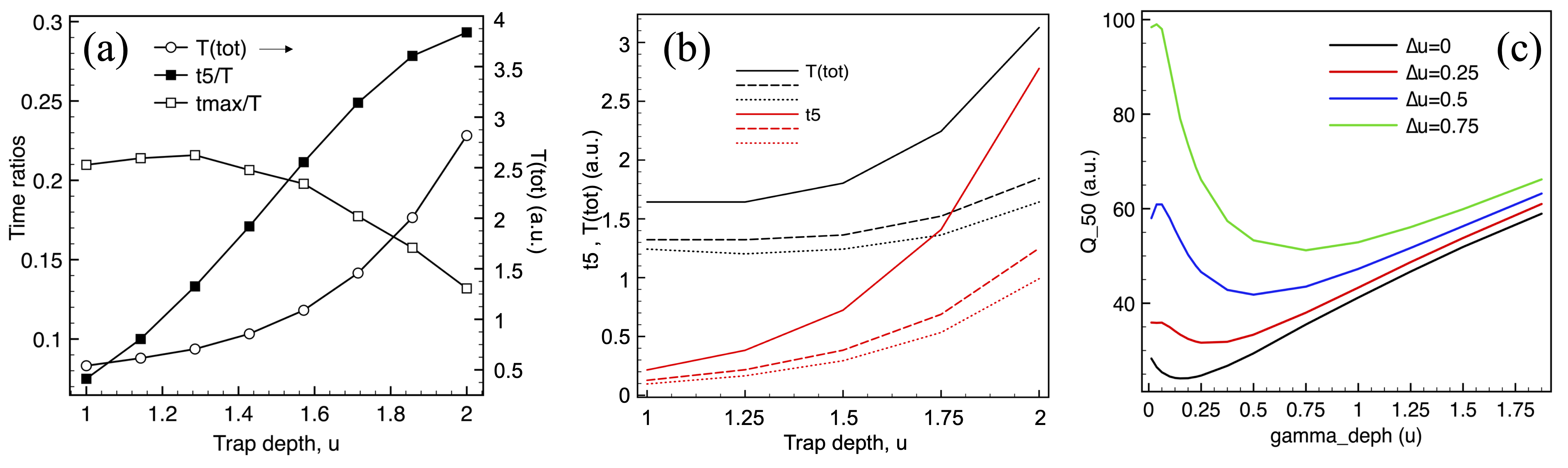}
\caption{\textbf{(a)} Effect of a single "trap" (bottleneck) at site 5 in the 8-site model with terminal sink. The total transit time $T_{tot}$ is shown on the right-side y-axis (open circles). Both the site-5 peak time $t_{max}/T$ (open squares) and the site-5 integral $t5/T$ (filled squares) are shown on the left-side y-axis. Trap depth is given in units of $u$, the common on-site energy (see text). \textbf{(b)} Effect of the superexchange mechanism between sites 4-6 when a trap is present on site 5. Total transit time = black plots; $t5$-residence (integral) time on site 5 = red plots. Full/dashed/dotted lines respectively correspond to $T_{4,6}=0,0.05u,0.1u$. \textbf{(c)} Evidence of the Environment-Assisted Quantum Transport in the 8-site model with alternating ("sawtooth") site energies $(u,\Delta)$, and relaxation constant $\Gamma=10^{-3} u$. The color plots correspond to $\Delta = 0, u/4, u/2, 3u/4$. A minimum of the total transfer time is observed for each curve, at increasing values of the dephasing constant $\gamma$.}
\label{fig5}
\end{figure*}

Here we manually introduce a "bottleneck" at a middle site 5 (ideally matching N5), by varying the onsite energy 
between $-u,...-2u$ (note that $-2u$ is the same value as the terminal sink site). Figure \ref{fig5}a shows the effect of increasing the "trap" energy on the electron transit times. Firstly, the total transit time to reach the terminal site is more than doubled (plot w/ open circles). The plot labelled "$t5/T$" (filled squares) represents the site residence time, integral of time spent by the electron on site 5, normalized by the total transit time. It is observed that this ratio increases more than fourfold, at the same time the absolute value of $t5$ is almost as large as the total transit time $T$ in the absence of any bottleneck. Also, the peak time of the N5 population, showed by the plot with open squares, is decreased by about half. Overall, the results show that upon increasing the depth of the "trap" site, the electron arrives earlier on the trap site and spends there the biggest part of the total transit time, which in turn is largely slowed down. 

\subsection{Super-exchange}\label{superex}

However, the proteins making up the electron transfer patch are not static blocks. Stuchebrukhov for example \cite{stuche1} suggested that small water molecules, stably trapped within the protein channels between N5 and N6a, could act as "superexchange bridges" that bypass the trap sites, boosting the electronic coupling by orders of magnitude and saving the system from total blockage. Several studies, for example using ruthenated proteins to modify the intersite distance \cite{beratan,winkler,winkler2}, and two-photon or 2D-IR ultrarapid spectroscopy \cite{hochstrass,romero}, have demonstrated the feasibility of superexchange pathways in both photosynthetic and respiratory protein complexes.

Here we can manually include super-exchange by introducing $T_{i,i\pm 2}$ terms with variable strength in the coupling matrix. For example, in the case of a trap of variable depth $(u,...2u)$ at site $i$=5, we add a $T_{4,6}=T_{6,4}=u/20$ or $u/10$, to observe the effect of alternate pathways sidestepping the trap site. Note that since the dependence of the hopping integrals $T_{ij}$ on the distance is exponential (with a factor linked to the Franck-Condon activation energy), doubling the $T$ may correspond to shortening the 4-6 intersite distance by more than just half the bare value. 

Figure \ref{fig5}b displays the effect of $T_{4,6}$ for $-\epsilon_5=u,...2u$ ("trap" depth). The plots respectively show the absolute values (a.u.) of total transit time (black) and $t5$ residence time (red), as a function of $u$ on site 5, with $T=0$ (full lines), $T=u/20$ (dashed lines) and $T=u/10$ (dotted lines). It is seen that $T=u/20$ already has a large effect, by reducing both the total and $t5$ times by large factors. Upon increasing further $T=u/10$, the transit time becomes nearly independent on the trap energy, except for a little increase at the last $-\epsilon=2u$. Also the $t5$ residence time becomes less dependent on the on-site energy value, at least at the lower range. 

While such results are only qualitative, they clearly demonstrate that the TEL model is able to capture both the botteneck effects of discontinuous energy profiles along the redox chain, and the possible effects of alternate transport pathways that may short-circuit such discontinuities. A better description of the actual pathways, also trying to distinguish between different transport modes (e.g., bridging waters, or aromatic vs aliphatic bridges), would certainly require to include explicit spatial effects in the description of the $T_{ij}$ hopping integrals.

\subsection{Environment-assisted quantum transport}

Electron transport in such biological systems seems to rest on an irreconcilable logical contradiction: how can a phenomenon be guided by a purely quantum law (tunneling), if quantum effects (coherence, phases, interferences) are rapidly and continuously destroyed by the protein thermal motions?
This is the apparent result of an accurate tuning that natural selection has refined over billions of years. A "dephasing-assisted transport" regime has been identified by the pioneering studies by Plenio \& Huelga \cite{plenio}, and by Rebentrost et al. \cite{reben}, who called the phenomenon \textit{Environment-Assisted Quantum Transport} (ENAQT). They demonstrated that the proton transfer efficiency of light-harvesting molecules such as the photosynthetic FMO complex, would be destroyed by both a pure quantum coherence and a classical full incoherence. Therefore, the system adjusts on an optimum compromise between the two opposite effects, for a condition that minimizes the charge transit time to a biologically relevant scale.

Here we will explore the ENAQT conditions, upon switching the on-site energy profile from a flat to a "sawtooth" one, as
$-\epsilon_i = [0,u,\Delta,u,\Delta,u,\Delta,u,2u]$, with $\Delta=u,...2u$. In the TEL model, the condition to observe such regime impose rather strict conditions on the relative coupling strength of the dephasing ($\gamma$) and relaxation 
($\Gamma$) terms.
In a simple model, the dephasing transfer rate scales as
\begin{equation}
k^{deph} \sim 2 T_{ij}^2 \frac{ \gamma }{ (\Delta^2 + \gamma^2) }
\label{hak}
\end{equation}
This is the standard Haken-Strobl adiabatic-elimination form \cite{hakenstro,nemoto}, which has a maximum when 
$\gamma \sim | \Delta |$.
So, dephasing helps when it broadens mismatched on-site energies enough to overcome thermal relaxation, but it hurts when it becomes too strong and produces the quantum-Zeno limit. In practice, dephasing-assisted regime or ENAQT is manifested between the two extremes: for $\gamma \ll | \Delta |$ we have
$k^{deph} \sim 2 T_{ij}^2 \gamma / \Delta^2$,
in which case dephasing helps, but for $\gamma \gg | \Delta |$ it is
$k^{deph} \sim 2 T_{ij}^2 / \gamma$,
and dephasing suppresses transport.

By comparing the dephasing and relaxation (forward/backward jump) terms, it can be observed that ENAQT would be visible only if the dephasing is not buried under the random jump contribution. 
Since the coherent term is $\sim T_{12}^2 / | \Delta |$  at the maximum $\gamma \sim | \Delta |$ of Eq.(\ref{hak}), a minimum visibility condition of the ENAQT effect is $\Gamma <  T_{ij}^2 / | \Delta |$. With the above choice of parameters, $T\sim u/20$, $\Delta \sim u$, and $u\sim 80$ meV for the typical energy scale of Complex I, we must set $\Gamma <  T_{ij}^2 / | \Delta | \sim 0.2$, while exploring  $\gamma$ values in a range of about [0.1 - 100] (that is, between $\sim 0$ and $\sim u$).

Figure \ref{fig5}c displays the plots of the Q50 time, that is the time necessary for the probability of filling the last (quinone) sink site is 50\%, as a function of $\gamma$, for $\Gamma=0.1$ (or 10$^{-3}u$) (We chose this time observable because it represents an integrated probability, instead of the total transit time $T$, however the two are related as $Q50\sim T\ln2)$.) Each color curve corresponds to a "sawtooth" distribution of the on-site energies, from flat ($\Delta=u$), to increasingly discontinuous ($\Delta>u$). A clear minimum of the transport time is observed for each curve, at values of $\gamma_{min}$ that increase nearly linearly with $\Delta/u$, and an amplitude (difference between the transport time at $\gamma=0$ and the time at $\gamma_{min}$) that increases accordingly.  This is a clear signature of dephasing-assisted regime, the system trying to optimize the shortest transit time in presence of combined dephasing and relaxation of the quantum coherence.

It may be noted that these results seem at odds with the findings of Dorner et al. \cite{dorner}, who found that the ENAQT effect is larger for a nearly flat on-site distribution, and tends to be reduced for a more rugged energy landscape. However, it should also be noted that energy profiles in that work were taken constant and the barrier disorder was characterized by varying the temperature, whereas here we use a constant temperature and modify the on-site energies, to model chemically different systems.

\section{Discussion}

Over the past decade, Lindblad master equations have become an increasingly attractive framework for modeling quantum transport in biological systems, because they combine computational efficiency with a direct quantum-mechanical description of dissipation. Nevertheless, the flexibility of the formalism also represents a potential weakness: different choices of jump operators can generate similar transport kinetics while leading to substantially different thermodynamic behavior. Taking as reference 
the common description of electron transport through biological redox chains,
in the present work we addressed these issues by comparing three complementary formulations of Lindblad dynamics: a conventional empirical model (EL), a thermodynamically consistent microscopic construction (BtL), and an intermediate empirical formulation enforcing detailed balance (TEL).  This comparison made it possible to distinguish which features of the dynamics are robust physical predictions, and which arise from the particular mathematical structure of the dissipator.

The three models examined in this work also provide three distinct physical pictures of electron transport. The EL formulation can be viewed as the quantum analogue of the classical Pauli master equation: electrons tend to remain localized on individual redox centers and are transferred sequentially through discrete, essentially irreversible jump events. Quantum coherence appears only as a correction superimposed on an otherwise classical hopping dynamics. At the opposite end, the BtL formulation describes transport as the evolution of a continuously distributed probability wave, where dissipation and thermal fluctuations emerge from a microscopic construction satisfying thermodynamic consistency by design. This representation naturally captures the collective interaction between the electronic state and the environment, although the corresponding position and momentum operators no longer possess a direct interpretation as localized observables on individual cofactors. The TEL appears to combine many of the advantages of both approaches. By enforcing local detailed balance through thermally dressed transition rates, it retains the intuitive site-resolved description of electron motion (notably in the current $\hat{J}$) while reproducing the thermodynamic behavior of the microscopic BtL formulation to a remarkable degree. The close agreement between these otherwise independent constructions suggests that the essential physics of biological electron transport is already captured by a thermodynamically-constrained Lindblad description, while the polaron interpretation provides a natural microscopic rationale for its applicability beyond the bare weak-coupling picture.

A central result of our study is that enforcing thermal detailed balance substantially modifies the physical interpretation of electron transport, without sacrificing the practical simplicity of the Lindblad approach. While the EL reproduces the expected directional transport, it may simultaneously generate heat and entropy exchanges incompatible with thermal equilibrium and produce coherent current oscillations that are best interpreted as artifacts of the chosen dissipator. By contrast, both the microscopic BtL formulation and the phenomenological TEL model satisfy the expected thermodynamic constraints while preserving the essential transport characteristics. The close quantitative agreement between the latter two independent approaches, perhaps the most surprising result of this work, strongly suggests that detailed balance constitutes the key ingredient required to obtain physically realistic weak-coupling descriptions of biological electron-transfer chains.

Beyond the methodological comparison, the present framework naturally reproduces several characteristic features of biological electron transport, including transient quantum coherence, bottleneck effects arising from heterogeneous redox landscapes, superexchange pathways, and environment-assisted quantum transport. These phenomena emerge within a unified open-system description that connects coherent tunneling, dissipative relaxation, and thermal fluctuations without abandoning complete positivity or computational tractability. The resulting formalism therefore provides a convenient platform for systematically exploring how protein architecture, energetic disorder, and environmental interactions jointly determine electron-transfer efficiency.

More generally, the results suggest that thermodynamic consistency should become a standard criterion when constructing Lindblad models for molecular and biological electron transport. Because the proposed formulation remains computationally inexpensive while incorporating physically motivated thermal constraints, it can readily be extended to realistic respiratory complexes, bacterial cytochrome networks, artificial molecular wires, and other nanoscale charge-transfer systems. We believe that the present work will contribute toward establishing thermodynamically grounded Lindblad approaches as practical tools for investigating non-equilibrium quantum transport in chemically and biologically relevant environments.


\begin{acknowledgments}
F. C. thanks the support of FY2026 JSPS/OF361 Invitational Fellowships for Research in Japan, and the kind hospitality of prof. Y. Sakai in the Chemical Systems Engineering Laboratory, The University of Tokyo, during an extended stay in the Spring 2026 term. Useful discussions about the biology of Complex I with dr. E. Leclerc (LIMMS and University of Tokyo) are gratefully acknowledged.  
\end{acknowledgments}

\bibliography{lindb3}

\clearpage

\appendix

\section{Numerical implementation}\label{workf}

In practice we do not calculate the true eigenstates (the \textit{adiabatic} basis) of the model Hamiltonian Eq.(\ref{hamil}) in order to parameterize the Lindblad equation. Instead, we must construct and solve the entire equation using a \textit{diabatic}-site basis.
 Because the cofactors are separated by relatively large distances (8-14~\AA) the electronic couplings ($T_{ij}$) are extremely weak, which makes the tight-binding approximation appropriate. Therefore, physical transport is fundamentally a non-adiabatic transition between distinct, localized chemical species.

\noindent \textbf{Hamiltonian Matrix.} To apply the Lindblad method in a numerical simulation imitating Mitochondrial Complex-I, in practice we define the diabatic orthonormal basis of 8 states, $\ket{1}, \ket{2} \dots \ket{8}$, where each state represents the electron fully sitting on the $i$-th cluster.
Instead of finding eigenstates from the Hamiltonian,  the matrix elements are estimate using separate quantum-chemistry and electrostatic methods. For the diagonal elements $\varepsilon_{i}$, typically Continuum Electrostatics (e.g., solving the Poisson-Boltzmann equation) combined with Molecular Dynamics (MD) has been used to find the standard redox potential of each cluster embedded in the protein matrix \cite{stuche1,stuche2}.
For the off-diagonal elements $T_{ij}$, other works used semi-empirical quantum methods (ZINDO or MNDO, again in tight-binding approximation) to calculate the tunnel coupling through the bridging amino acids, directly in the localized state representation (e.g. \cite{castellano}).

\noindent \textbf{Localized Lindblad Operators.} Since we are already in the site basis, the pure dephasing operators are simply the standard projection matrices:
\begin{equation}
\hat{L}_i = \ketbra{i}{i}, 
\end{equation}
This allows to either use a unique, average value for all cofactors (mono-channel mode), or to assign a unique dephasing rate $\gamma _{i}$ to each individual cofactor (multi-channel mode), based on the features of each protein and its immediate environment.

\noindent \textbf{Numerical Integration.} With $\hat{H}_{electronic}$ and all $\hat{L}_{k}$ explicitly written as simple 8$\times$8 matrices in the site basis, we feed them directly into an open-source quantum dynamics solver from the QuTiP python library \cite{qutip1,qutip2}. The \textit{mesolver()} master equation solver integrates the differential equations in time using standard Runge-Kutta routines, without ever needing to diagonalize the Hamiltonian.

To avoid cluttering the script with complex conversions (such as Planck's constant scaling factors, like $\hbar = 6.582 \times 10^{-13}$ meV $\cdot$s), the time grid e.g. \textit{times = np.linspace}(0, 15, 300) is programmed in arbitrary scaled units. In a production script, this is explicitly scaled so that the entire transfer finishes in the physiological window of roughly 10 to 50 microseconds.
Coefficients are expressed as energies (meV) rather than frequencies (s$^{-1}$ or ps$^{-1}$) since the Lindblad equation solver in QuTiP natively works in a unit system with $\hbar = 1$. To go back from such energies to real times, the reduced Planck constant is used:  1 meV $\approx$ 1.52 ps$^{-1}$.

Typical values of the empirical Lindblad constants should be in the range $\gamma \simeq 2$ meV, suggested by the typical decoherence time $T_2^* =1/\gamma \sim 0.3$ ps; 
$\Gamma \simeq 1.2$ meV coming from Marcus' theory; in refs.\cite{stuche2,moser}, the jump rate between iron-sulphur clusters separated by about 10 \AA~(with water molecules acting as mediators) was estimated $(1.5 - 2.0) \times 10^{12}$ Hz, so that $\Gamma =\hbar \, (1.8\,10^{12}) = 1.2$ meV.

We used for both models (empirical vs. BtL) values of site energies with either a smooth profile, [0.0, -90.0, -90.0, -80.0, -70.0, -70.0, -70.0, -190.0], or a more pronounced "sawtooth" profile [0.0, -90.0, -90.0, -80.0, -70.0, -70.0, -70.0, -190.0], and
 $T_{ij}$ tight-binding couplings [5.0, 3.5, 2.0, 2.5, 3.0, 2.8, 4.0] \cite{stuche1,stuche2}. In the mono-channel mode (see below) 
 $\gamma$ = 2 meV and $\Gamma$ = 1.2 meV approximately reproduce physiological transit times for the empirical model. In the BtL mono-channel model, $m = 0.05$ (acts as a time-scale adjustment), $\beta = 0.45$ (meV s)$^{-1}$, $\omega$ = 0.632/$\hbar$ meV$^{-1}$ = 0.963 THz.  Temperature is 26.7 meV (310 K) and $\hbar$ = 1 in all cases. The relaxation time is adjusted so as to reproduce the $\hr_{88}(t)$ discharge curve of N2 from the empirical model, which is in turn adjusted on experimental data.

With the above values of redox potentials and $\hbar$=0.658 meV$\cdot$[s, the arbitrary time units shown in the plots are in the range of picoseconds. This is not physically unreasonable, however experimental values of electron transit times are in the 100-200 $\mu$s. This means that physiological rates are not rate-limited by the electron transfer but by other processes, e.g., slow injection, gating, Franck-Condon factors, conformational/proton-coupled steps, protein allotropic transformations.

\subsection{Example with two sites}\label{twosite}

As a bookkeping example to appreciate the competition between the quantum coherence and thermal dephasing and relaxation terms (described by the coupling constants $\gamma$ and $\Gamma$, respectively), let us take two adjacent redox sites labelled 1 and 2. Considering the "descending" flow of electrons from the site 1 (donor) to the site 2 (acceptor), the equations for each term of the density matrix are easily obtained:
\begin{align}
\frac{d\rho_{11}}{dt} &= -2 \, \textrm{Im}(T_{12} \rho_{12}) - \Gamma \rho_{11} \nonumber \\
\frac{d\rho_{22}}{dt} &= +2 \, \textrm{Im}(T_{12} \rho_{12}) + \Gamma \rho_{11} \nonumber \\
\frac{d\rho_{12}}{dt} &= 
-i (\varepsilon_1 - \varepsilon_2) \rho_{12} -i T_{12} (\rho_{22} - \rho_{11}) - \nonumber \\
& \phantom{aa} - \frac{(\gamma_1 + \gamma_2)}{2}\rho_{12} -
\frac{\Gamma}{2}\rho_{12} 
\label{twosta}
\end{align}
 The diagonal populations $\rho_{11}$ and $\rho_{22}$ measure the probability of finding electrons on either site. 
 The dissipation-relaxation term $\Gamma \rho_{11}$ on the diagonals describes the voiding of site 1 and filling of site 2.
 The off-diagonal terms measure the quantum superposition and phase stability between the two sites. 
 The $(\varepsilon_1 - \varepsilon_2)$ is a fast phase rotation that tends to destroy the coherence. 
 The imaginary term proportional to $(\rho_{22} - \rho_{11})$ produces quantum coherence as long as there is difference between the two populations.

 The imaginary tunneling diagonal term arises from the phase of the wavefunctions, in fact the explicit commutator from Eq.(\ref{tbeq1},\ref{tbeq2}) is:
 \begin{align}
 \bra{1} [\hat{H},\hr] \ket{1} &= \sum_{k=1}^2 H_{1k} \rho_{k1} - \sum_{k=1}^2 \rho_{1k} H_{k1} = \nonumber \\
 &= \varepsilon_{11} \rho_{11} + T_{12} \rho_{21} + \rho_{11} \varepsilon_{11} +  \rho_{12}  T_{21} = \nonumber \\
 & = 0 + T_{12} ( \rho_{21} - \rho_{12} )
 \label{tunnel1}
 \end{align}
for real coefficients $T_{12}=T_{21}$.  Evidently, if the density matrix were just real (no phase difference) there would be no tunneling at all, we are back to the classical equation. Instead $\rho_{21}=\rho_{12}^*$, meaning $\rho_{12}=x+iy$ and $\rho_{21}=x-iy$. Therefore, $\rho_{21} - \rho_{12} = -2iy = -2i \, \textrm{Im}(\rho_{12})$.

It can be noticed that even if we start at $t=0$ with a pure state, real density matrix:
\begin{equation}
\hr(0) = \left(
\begin{matrix}1&0\\ 0&0\end{matrix} 
\right)
\end{equation}
the off-diagonal terms with a complex value of the derivative immediately create quantum coherence and activate the tunneling:
\begin{equation}
\frac{d\rho_{12}}{dt} = -i \, T_{12} (\rho_{22} - \rho_{11}) = -i \, T_{12} (0 - 1) = +i \, T_{12}
\end{equation}

For the off-diagonal term the tunneling is:
 \begin{align}
 \bra{1} [\hat{H},\hr] \ket{2} &= -i \left( \sum_{k=1}^2 H_{1k} \rho_{k2} - \sum_{k=1}^2 \rho_{1k} H_{k2} \right) = \nonumber \\
 &= -i \left( \varepsilon_{11} \rho_{12} + T_{12} \rho_{22} - \rho_{11} T_{12} - \rho_{12}  \varepsilon_{22} \right) = \nonumber \\
 & = -i (\varepsilon_{22} - \varepsilon_{11}) \rho_{12} -i  ( \rho_{22} - \rho_{11} )T_{12}
 \label{tunnel2}
 \end{align}


\section{Implementation of the BtL model for a redox chain}\label{giordx}

The Lindblad operators can be recast in the form of the BtL paper \cite{giordan}, with $\hq,\hp$ operators of the protein, $\hbar \omega$ the average phonon frequency, and a common position-momentum dissipation factor $\beta$ for which the condition $\beta_{p}=m^{2}\omega ^{2}\beta_{q}=\beta$, is enforced:

\begin{align}
\frac{d\rho}{dt} &= \frac{1}{i\hbar} [\mathcal{H}_0,\rho] 
- \frac{k_BT \beta m}{\hbar^2} [q,[q,\rho]] + 
  \frac{k_BT \beta}{m(\hbar\omega)^2} [p,[p,\rho]] + \nonumber \\
  & + \frac{\beta}{2i\hbar} \Theta(\omega,T) \Big( [q,\{p,\rho\}] - [p,\{q,\rho\}] \Big)  
\label{giord}
\end{align}
The factor $\Theta(\omega,T) =\tanh \xi  / \xi$ with $\xi=\hbar \omega / 2k_BT$,
is the explicit form of the friction operators $\Theta$ in the paper, reduced to the on-site oscillator model. The "mass" is not the electron mass, but an effective thermal inertia symbolically enveloping the protein vibrational degrees of freedom.

The three last terms at RHS in Eq.(\ref{giord}), however, do not correspond one-to-one to the purely dephasing and relaxation terms in the empirical Lindblad formulation above (RHS of Eq.(\ref{gskl})). They actually mix the two processes in a symmetrical  way, although the physical meaning is different.
For example, the first term contains double commutators like $[q,[q,\rho]]$. By expanding the term:
\begin{equation}
[q,[q,\rho]] = q^2\rho +\rho q^2 - 2q\rho q = \{q^2,\rho\} - 2q\rho q,
\end{equation}
that is, an anticommutator contributing to the relaxation part, and a "sandwich" term contributing to the dephasing term of the Lindblad dissipator. Moreover, these terms are multiplied by the temperature, suggesting the increasing role of dephasing and relaxation at physiological conditions. Also the last term of Eq.(\ref{giord}) shows this mixing of anticommutator and sandwich terms, in this case multiplied by the scaling correction $Q(\omega,T)$: this factor (related to the hyperbolic tangent) is a purely quantum correction ensuring that energy loss occurs in accordance with the uncertainty principle, preventing the system from "dissipating" below the quantum ground state.

In the empirical/phenomenological EL model of Section \ref{workf} taking dephasing 
$\gamma$-terms $\vert 1 \rangle \langle 1 \vert$ and relaxation $\Gamma$-terms $\vert 2 \rangle \langle 1 \vert$ as completely separate things is a physical shortcut. The basic formula of BtL paper \cite{giordan} shows that in physical reality, noise and dissipation are inextricably linked. In order to recover a completely positive evolution, the position operator $q$ and the momentum operator $p$ must enter both the thermal dephasing block (second term of (\ref{giord})) and the dissipative relaxation block (third term).

\subsection{Mapping $q$ and $p$ on the tunneling sites}

In a tight-binding formulation of the harmonic oscillator, we do not model the electron as a particle that physically oscillates in free space like a spring. Instead, the flow of charge in the discrete lattice of $N$ sites is reproduced using the algebra of creation/annihilation operators, which shares the exact same mathematical structure as the harmonic oscillator despite the different physical meaning.

We introduce the  second-quantization of the position $\hq$ and the momentum $\hp$ by the raising and lowering operators $\cd,\cc$. For the chain of $N$ cofactors these operators must become discrete spatial "jump operators":
\begin{itemize}
\item The destruction operator $\cc$ becomes the operator that makes the electron jump forward along the chain, emptying the previous site, as $\sum_i \vert i+1 \rangle \langle i \vert$. 
\item The creation operator $\cd$ becomes the jump backward (or population creation on the site) as $\sum_i \vert i \rangle \langle i+1 \vert$.
\end{itemize}
Consequently, the $\hq$ and $\hp$ operators entering Eq.(\ref{giord}) convert to explicit, symmetric $N \times N$ matrices:
\begin{align}
\hq&= \sqrt{\tfrac{\hbar }{2m\omega }}(\cc + \cd) \nonumber \\
\hp&= -i \sqrt{\tfrac{\hbar m\omega }{2}}(\cc - \cd)
\end{align}
from which the following matrix operators (Lindblad-dephasing-relaxation combined) can be constructed: 
\begin{align}
\hat{{\mathbf{Q}}}_{\text{Complex\ I}} &=  \phantom{aa} \sqrt{\tfrac{\hbar }{2m\omega }} \sum_{i=1}^{7} \Big( |i+1\rangle \langle i|+|i\rangle \langle i+1| \Big) \nonumber \\
\hat{{\mathbf{P}}}_{\text{Complex\ I}}  &= -i \sqrt{\tfrac{\hbar m\omega }{2}} \sum_{i=1}^{7} \Big( |i+1\rangle \langle i|-|i\rangle \langle i+1| \Big)
\label{gem1}
\end{align}

The density operator is the expectation value of the site occupation, $\hr_ij = \langle \cd_i\cc_j \rangle$, the off-diagonal terms representing transitions between sites. 

The matrix $\hat{\mathbf{Q}}$ is a purely real and symmetric tridiagonal matrix. It represents the spatial coordinate discretized along the quantum wire. The matrix $\hat{\mathbf{P}}$ is a purely imaginary and antisymmetric tridiagonal matrix. It represents the native quantum current or transition momentum between adjacent clusters.

The Lindblad operators can be conveniently cast in the form:
\begin{align}
 \hat{L}_1 &= a \, \hq \,\, + \,\, i \,b \hp  \nonumber \\
 \hat{L}_2 &= c \, \hq \,\, + \,\, i \, d \hp
\label{gem3}
\end{align}

While an even more general choice would be to use complex coefficients for all the four terms, in practice the choice of real values for the $\hq$ part and purely imaginary values for the $\hp$ part remains physically motivated, and restrains the search space. In particular, it forces the condition $B=-D$, which is important to retrieve the particular form of Eq.(\ref{giord}), see next section \ref{check}. 

For $\hat{L}_1,\hat{L}_2$ the explicit coefficients $a,b,c,d$ can be obtained from the Kossakowski matrix construction. Using the following simplification in Eq.(\ref{giord}):
\begin{equation}
A = \frac{k_BT \beta m}{\hbar^2}, \phantom{a} C = \frac{k_BT \beta}{m(\hbar\omega)^2}, \phantom{a} B = -D = \frac{\beta}{2i\hbar} \Theta(\omega,T), 
\label{ABCDdef}
\end{equation}
a Kossakowski matrix is constructed as:
\begin{equation}
\bf{K} = \left( \begin{matrix} -2A & -i(B-D)/\hbar \\ +i(B-D)/\hbar & -2C \end{matrix} \right)
\label{kossa}
\end{equation}
whose eigenvalues and eigenvectors are found as:
\begin{align*}
\lambda_{1,2} &= - (A+C) \pm 2 \sqrt{(A+C)^2 + \frac{(B-D)^2}{\hbar^2} } \phantom{aaaa} (\pm\textrm{ for 1,2})   \\
\\
\textbf{v}_1 &= \left( \begin{matrix} \cos \theta \\ i \sin \theta \end{matrix} \right),\phantom{a}
\textbf{v}_2 = \left( \begin{matrix} -\sin \theta \\ i \cos \theta \end{matrix} \right),\phantom{a}
\theta = \frac{1}{2} \arctan \left( \tfrac{B-D}{\hbar(C-A)} \right)
\end{align*}

The spectral decomposition $\textbf{W} = \textbf{V}\sqrt{\Lambda}$ gives the coefficients of the Lindblad operators as the columns of $\textbf{W}$ (with the choice $\mu=0$, it is $B=-D$:
\begin{align}
\hat{L}_1 &= \sqrt{\lambda_1} \, (\cos\theta \, \hq + i \, \sin\theta \, \hp)  \nonumber \\
\hat{L}_2 &= \sqrt{\lambda_2} \, (-\sin\theta \, \hq + i \, \cos\theta \, \hp)  
\end{align}

\subsection{Coherence of the Lindblad operators with the BtL equation}\label{check}

We can now check that the solver \textit{mesolve()}, which blindly computes the sum of Lindblad dissipator for all the elements of a list (called \textit{c\_ops} in the code), in this case just the sum of $L_1,L_2$, gives back the exact equation of the BtL model:
\begin{align}
\mathcal{D}(\hr) &= \left( \hat{L}_1\hr\hat{L}_1^{\dag} - \frac{1}{2} \{\hat{L}_1^{\dag}\hat{L}_1,\hr \} \right)   + \nonumber \\
& \phantom{aaaaaaaa} + \left( \hat{L}_2\hr\hat{L}_2^{\dag} - \frac{1}{2} \{\hat{L}_2^{\dag}\hat{L}_2,\hr \} \right)
\end{align}

If we now make the explicit substitution of $\hq, \hp$ and develop separately the two original terms "sandwich" and anticommutator of Eq.(\ref{gskl}):

\begin{align}
 \hat{L}_1\hr\hat{L}_1^{\dag} &= ( a \hq + i \, b\hp)\hr ( a\hq - i \, b \hp) = \nonumber \\
 &=  a^{2} \hq\hr\hq + b^2 \hp\hr\hp  +  i \, ab ( \hp\hr\hq - \hq\hr\hp )  \\
 \hat{L}_2\hr\hat{L}_2^{\dag} &= ( c \hq + i \, d\hp)\hr ( c\hq - i \, d \hp) = \nonumber \\
 &= c^{2} \hq\hr\hq + d^2 \hp\hr\hp  +  i \, cd ( \hp\hr\hq - \hq\hr\hp ) \\
\{\hat{L}_1^{\dag}\hat{L}_1,\rho\} &= \{ ( a\hq - i \, b \hp) ( a \hq + i \, b\hp), \hr \} = \{ a^2 (\hq\hq\hr + \hr\hq\hq )+ \nonumber \\
 & \phantom{aa} + b^2 (\hp\hp\hr + \hr\hp\hp )
 i\, ab(\hq\hp - \hp\hq) ], \rho \} = \nonumber \\
&= a^2 ( \hq\hq\hr + \hr\hq\hq ) + b^2 ( \hp\hp\hr + \hr\hp\hp ) + \nonumber \\
& \phantom{aa}+ i \, ab ( \hq\hp\hr + \hr\hq\hp -  \hp\hq\hr - \hr\hp\hq )  \\
\{\hat{L}_2^{\dag}\hat{L}_2,\rho\} &= \{ ( c \hq - i \, d \hp) ( c \hq + i \, d \hp), \hr \} = \{ c^2 (\hq\hq\hr + \hr\hq\hq )+ \nonumber \\
 & \phantom{aa} + d^2 (\hp\hp\hr + \hr\hp\hp )
 i\, cd(\hq\hp - \hp\hq) ], \rho \} = \nonumber \\
&= c^2 ( \hq\hq\hr + \hr\hq\hq ) + d^2 ( \hp\hp\hr + \hr\hp\hp ) + \nonumber \\
& \phantom{aa}i \, cd ( \hq\hp\hr + \hr\hq\hp -  \hp\hq\hr - \hr\hp\hq )
\end{align}
Now regroup the real and imaginary parts (including the -1/2 for the anticommutators):
\begin{align}
\textrm{Re} &= (a^2+c^2) ( \hq\hr\hq -\tfrac{1}{2}  \hq\hq\hr - \tfrac{1}{2} \hr\hq\hq ) + \nonumber \\
&+ (b^2+d^2) ( \hp\hr\hp -\tfrac{1}{2}  \hp\hp\hr - \tfrac{1}{2} \hr\hp\hp ) \\
\textrm{Im} &= (ab + cd) [ (\hp\hr\hq - \tfrac{1}{2} \hq\hp\hr + \tfrac{1}{2} \hp\hq\hr) - \nonumber \\
&- (\hq\hr\hp + \tfrac{1}{2} \hr\hq\hp - \tfrac{1}{2} \hr\hp\hq ) ]
\end{align}

In the real part we have terms like $(\hq\hr\hq - \hq\hq\hr/2 - \hr\hq\hq/2) = [\hq,[\hq,\hr]]/2$, and same for the $\hp$ terms, so that:
\begin{equation}
\text{Re} =  -\frac{a^2+c^2}{2} [\hq,[\hq,\hr]] - \frac{b^2+d^2}{2} [\hp,[\hp,\hr]]
\end{equation}

For the imaginary part, it is readily verified that:
\begin{align}
&-\frac{1}{2} \Big( [ \hq,\{\hp,\hr\} ] - [\hp,\{\hq,\hr\}] \Big) = \nonumber \\ 
& = -\frac{1}{2} \Big( \hq,(\hp\hr + \hr\hp)] - \frac{1}{2} [ \hp,(\hq\hr + \hr\hq) \Big) = \nonumber \\
& = - \frac{1}{2} \Big( \hq (\hp\hr + \hr\hp) - (\hp\hr+\hr\hp)\hq - \hp(\hq\hr+\hr\hq) + \nonumber \\
& \phantom{aa}+ (\hq\hr + \hr\hq) \hp \Big) = \nonumber \\
& =(\hp\hr\hq - \tfrac{1}{2} \hq\hp\hr + \tfrac{1}{2} \hp\hq\hr) 
 - (\hq\hr\hp + \tfrac{1}{2} \hr\hq\hp - \tfrac{1}{2} \hr\hp\hq )
\end{align}
that is, exactly the term in $[...]$ for the imaginary part above. Therefore, regrouping the two terms, we finally have:
\begin{align}
\mathcal{D}(\hr)
&= -\frac{(a^2+c^2)}{2} [\hq,[\hq,\hr]] - \frac{(b^2+d^2)}{2} [\hp,[\hp,\hr]] \nonumber \\
& \phantom{aa}+ \frac{(ab+cd)}{2i} \Big( [ \hq,\{\hp,\hr\} ] - [\hp,\{\hq,\hr\}] \Big)
\end{align}

With the right coefficients $A,B,C,D$, this is just Eq.(\ref{giord}) of the BtL model.

\end{document}